\makeatletter
\declare@file@substitution{revtex4-1.cls}{revtex4-2.cls}
\makeatother
\documentclass[twocolumn]{aastex631}
\usepackage{newtxtext,newtxmath}
\usepackage[T1]{fontenc}
\usepackage{CJKutf8}
\usepackage{ae,aecompl}
\usepackage{graphicx}      
\usepackage{amsmath}
\usepackage{amssymb}	
\usepackage{float}
\usepackage[]{hyperref}

\usepackage{booktabs}
\usepackage{tablefootnote}
\usepackage{threeparttable}
\usepackage{natbib}
\usepackage{hyperref}
\usepackage{graphicx}
\definecolor{darkblue}{rgb}{0.6,0.0,0.0}
\definecolor{cyan}{rgb}{0.0,0.0,0.6}

\received{}
\revised{}
\accepted{}

\begin{document}

\title{Unraveling Chemical Enrichment in Extreme Emission-Line Galaxies: A Multi-Element Bayesian View of Bursty Star Formation and Galaxy Evolution in DESI}

\date{April 2026}

\shorttitle{Multi-Element Enrichment in EELGs}
\shortauthors{Razieh Emami et. al.}

\correspondingauthor{Razieh Emami}
\email{razieh.emami$_{-}$meibody@cfa.harvard.edu}

\author[0000-0002-2791-5011]{Razieh Emami}
\affiliation{Center for Astrophysics $\vert$ Harvard \& Smithsonian, 60 Garden Street, Cambridge, MA 02138, USA}

\author[0000-0002-9081-2111]{James A.\@ A.\@ Trussler}
\affiliation{Center for Astrophysics $\vert$ Harvard \& Smithsonian, 60 Garden Street, Cambridge, MA 02138, USA}

\author[0000-0003-4512-8705]{Tiger Yu-Yang Hsiao}
\affiliation{Department of Astronomy, The University of Texas at Austin, Austin, TX 78712, USA}
\affiliation{Cosmic Frontier Center, The University of Texas at Austin, Austin, TX 78712, USA}

\author[0000-0002-8810-858X]{Kaley Brauer}
\affiliation{Center for Astrophysics $\vert$ Harvard \& Smithsonian, 60 Garden Street, Cambridge, MA 02138, USA}

\author[0000-0001-6950-1629]{Lars Hernquist}
\affiliation{Center for Astrophysics $\vert$ Harvard \& Smithsonian, 60 Garden Street, Cambridge, MA 02138, USA}

\author[0000-0003-4284-4167]{Randall Smith}
\affiliation{Center for Astrophysics $\vert$ Harvard \& Smithsonian, 60 Garden Street, Cambridge, MA 02138, USA}

\author[0000-0003-2808-275X]{Douglas Finkbeiner}
\affiliation{Center for Astrophysics $\vert$ Harvard \& Smithsonian, 60 Garden Street, Cambridge, MA 02138, USA}

\author[0000-0002-4622-6617]{Fengwu Sun}
\affiliation{Center for Astrophysics $|$ Harvard \& Smithsonian, 60 Garden St., Cambridge, MA 02138, USA}

\author[0000-0002-3324-4824]{Rebecca Davies}
\affiliation{Centre for Astrophysics and Supercomputing, Swinburne University of Technology, Hawthorn, Victoria, Australia} 
\affiliation{ARC Centre of Excellence for All Sky Astrophysics in 3 Dimensions (ASTRO 3D), Australia}

\author[0000-0002-5872-6061]{James F. Steiner}
\affiliation{Center for Astrophysics $\vert$ Harvard \& Smithsonian, 60 Garden Street, Cambridge, MA 02138, USA}
\author[0000-0001-8593-7692]{Mark Vogelsberger}
\affil{Department of Physics and Kavli Institute for Astrophysics and Space Research, Massachusetts Institute of Technology, Cambridge, MA 02139, USA}
\affil{Fachbereich Physik, Philipps Universit\"at Marburg, D-35032 Marburg, Germany}

\author[0000-0002-3642-2446]{Tobias Looser}
\affiliation{Center for Astrophysics $\vert$ Harvard \& Smithsonian, 60 Garden Street, Cambridge, MA 02138, USA}

\author[0000-0001-6950-1629]{Grant Tremblay}
\affiliation{Center for Astrophysics $\vert$ Harvard \& Smithsonian, 60 Garden Street, Cambridge, MA 02138, USA}

\author{Letizia Bugiani}
\affiliation{Dipartimento di Fisica e Astronomia, Universit`a di Bologna, Bologna, Italy}

\begin{abstract}
Extreme emission-line galaxies (EELGs) provide a sensitive probe of chemical enrichment in low-mass galaxies undergoing bursty star formation, where the interplay between star formation, feedback, and gas accretion remains poorly constrained. Using DESI DR1, we select 23 nearby EELGs with simultaneous detections of 19 ionic species ($\mathrm{S/N}\geq 4$), stellar masses $M_*\geq 10^{7},M_{\odot}$, and extreme equivalent widths in H$\alpha$ and [O~III]~$\lambda5007$ ($\mathrm{EW}\geq 500,\mathrm{\AA}$). We infer non-parametric star-formation histories with \texttt{BAGPIPES} and fit a multi-element Bayesian single-zone chemical-evolution model to O, N, Ne, S, and Ar abundance ratios, allowing time-dependent star-formation efficiency, outflow mass loading, and an inflow metallicity that evolves from primordial to recycled gas.
We find short present-day depletion timescales and large mass-loading factors, implying rapid gas cycling in a burst-driven, non-equilibrium regime, with depletion times systematically below expectations from the Kennicutt–Schmidt relation. While star-formation efficiency and outflow parameters are robust across modeling assumptions, the inflow metallicity remains more weakly constrained, reflecting degeneracies between metal production and gas accretion.
The differential behavior of abundance ratios separates the dominant physical drivers: star-formation efficiency governs the evolutionary progression in abundance space, outflows regulate metal retention and the normalization of $\mathrm{X/O}$, and inflow metallicity sets the baseline enrichment. Among these, $\mathrm{N/O}$ provides strong leverage on burst timing and gas flows, $\mathrm{Ne/O}$ remains nearly invariant, and $\mathrm{S/O}$ and $\mathrm{Ar/O}$ exhibit intermediate sensitivity.
These results demonstrate that multi-element abundance patterns enable a direct and physically interpretable reconstruction of baryon-cycle processes in extreme low-mass starbursts.

\end{abstract}


\section{Introduction}
\label{sec:Intro}

Dwarf galaxies with present-day stellar masses $M \leq 10^9 ~\mathrm{M}_{\odot}$ represent a large population of galaxies in the universe and exhibit remarkable diversity in their morphologies, star-formation histories, and environments \citep{1957moas.book.....Z,1971ARA&A...9...35H, 1992ARA&A..30..613K,2004AJ....127.2031K,2019ApJ...874...93B}. Owing to their relatively low mass and high number densities, dwarf galaxies serve as critical testbeds for models of galaxy formation and evolution across cosmic time. In particular, their low stellar masses render them ideal probes of feedback-regulated star formation, environmental quenching, and baryonic processes at the low-mass end of the galaxy population. Moreover, within the framework of hierarchical structure formation, these systems are thought to constitute the primordial building blocks of more massive galaxies \citep{2011ApJ...739....5W}, offering direct insights into the earliest phases of galaxy assembly. 

This theoretical perspective is now being richly tested by recent JWST observations, which have revealed a population of extremely distant, low-mass galaxies at $z \geq 10$
\citep{2006SSRv..123..485G,2023PASP..135f8001G,2023arXiv231012340E,2023ApJS..269...16R,2023ApJ...946L..13F,2023ApJS..265....5H,2023NatAs...7..622C,2023ApJ...957L..34W,2024ApJ...970...31R,2024Natur.633..318C,2024ApJ...971...43M,2024Natur.633..318C,2025arXiv250311457N,2025ApJ...980..225T,2025arXiv250100984W,2025A&A...696A..87C}. Dwarf galaxies are also widely considered as key contributors to the universe's ionizing photon budget during the epoch of reionization \citep{2008MNRAS.385L..58C,2014MNRAS.437L..26S,2015ApJ...802L..19R,2015ApJ...811..140B,2015ApJ...814...69A}. Despite their importance, significant uncertainties remain in characterizing their star-formation histories (SFHs) as well as their chemical enrichment pathways. 

Theoretically,  recent hydrodynamical simulations predict that star formation in low-mass galaxies is highly bursty, driven by the interplay between stellar feedback and shallow gravitational potentials \citep{2004AA...422...55P, 2007ApJ...667..170S, 2014ApJ...792...99S}.

Observationally, galactic archaeology has provided crucial insights into the evolutionary histories of dwarf galaxies by reconstructing the ages of their stellar populations \citep{2009ARA&A..47..371T,2014ApJ...789..147W}. For instance, \citet{2011ApJ...742..111V} demonstrated that SFHs in dwarf galaxies are often complex and highly bursty, characterized by episodic star-forming events that produce on the order of $\sim 10^8~\mathrm{M}_{\odot}$ stars per burst \citep{1997RvMA...10...29G,1998ARA&A..36..435M,2011ApJ...739....5W}. Each episode results in the formation of massive stars, which synthesize heavy elements via stellar nucleosynthesis. Through stellar winds and supernova explosions, these nucleosynthetic products are expelled into the interstellar medium (ISM), where they mix with ambient gas and inflowing material. This enrichment process raises the gas-phase metallicity, which is then imprinted on subsequent generations of stars. As a result, the gas-phase metal abundance in dwarf galaxies is highly sensitive to their past star formation activity, the balance of gas inflows and outflows, and the efficiency of mixing processes within the ISM \citep{1988A&A...197...47T,1995A&A...294..432R,2019ApJ...874...93B,2024ApJ...964L..12R}.
Indeed, chemical enrichment in galaxies occurs in an open-box framework, where abundance patterns are influenced not only by stellar nucleosynthesis but also by the loss of metals through star formation-driven outflows \citep{1985MNRAS.217..391M}.
The buildup of heavy elements through stellar nucleosynthesis is therefore directly linked to galaxy evolution and serves as a key tracer of the growth and evolutionary history of dwarf galaxies.

Thanks to recent advances in deep-field observations and the availability of high-resolution spectra from large spectroscopic surveys \citep{2001MNRAS.328.1039C,2005A&A...439..845L,2007ApJS..172...70L,2009MNRAS.399..683J,2010MNRAS.401.1429D,2011MNRAS.413..971D,2011AJ....142...72E,2011ApJ...741....8C,2012ApJS..200...13B,2013AJ....145...10D,2014A&A...566A.108G,2015A&A...576A..79L,2015ApJS..218...15K,2015ApJS..220...12S,2016ApJS..223...29V,2016AJ....151...44D,2017AJ....154...28B,2017A&A...608A...1B,2024AJ....167...62D}, we now possess unprecedented datasets that enable direct and robust measurements of gas-phase metallicities \citep{2013ApJ...765..140A,2019ApJ...874..125P,2020MNRAS.491.1427S}. These data provide critical constraints for the construction of models of galactic chemical enrichment (GCE) and the tracking of the buildup of metals over cosmic time \citep{1980FCPh....5..287T,1997nceg.book.....P,1997ApJ...477..765C,1999MNRAS.307..857B,2020MNRAS.498.1364J,2020ApJ...900..179K,2021MNRAS.508.4484J,2021A&ARv..29....5M}.

Detailed modeling of broadband spectral energy distributions (SEDs) allows for robust estimation of stellar masses and mass-weighted ages, along with constraints on star-formation rates, dust content and chemical composition.
%%%%%%%%%%%%%%%%%%%%%%%%%%%%%%%%%%%%%%%%%%%%%%%%%%%%
\begin{table*} [!htbp] 
\centering
\label{ion_list}
	\begin{tabular}{|l|c|r|} 
		\hline 
		Element &  Ion List
        \\ \hline
        {\color{blue} Hydrogen}  & H$\beta$ ; H$\alpha$
        \\ \hline
		{\color{blue} Oxygen}  & [O II] 3726 ; [O II] 3729 ;  [O II] 7320 ; [O II]7331 ; [O III] 4363 ; [O III] 4959 ; [O III] 5007
        \\    
        \hline
        {\color{blue} Sulfur}  & [S II] 6716 ; [S II] 6731 ; [S III] 6312 ;  [S III] 9069 ; [S III] 9533  \\ \hline
         {\color{blue} Argon}  &  [Ar III] 7136 ; [Ar III] 7751
        \\ \hline
        {\color{blue} Nitrogen}  &  [N II] 6583  \\ \hline
        {\color{blue} Neon}  & [Ne III] 3869 ; [Ne III] 3967 \\
        \hline
       	\end{tabular}
	\caption{The list of 19 ions used to derive the element abundance ratios in our galaxy sample from DESI DR1. }
\label{Tab:ion-list1}
\end{table*}
%%%%%%%%%%%%%%%%%%%%%%%%%%%%%%%%%%%%%%%%%%%%%%%%%%%%

The spectrum of Star-forming galaxies show strong emission lines \citep{1970ApJ...160..405S}, which provide a powerful diagnostic for measuring gas-phase metallicities and elemental abundances. Such systems are often undergoing intense bursts of star formation, characterized by specific star formation rates (sSFRs) of $\simeq 10\mathrm{Gyr}^{-1}$. 

Extreme emission-line galaxies (EELGs) \citep{2011ApJ...743..121A, 2014ApJ...791...17M, 2020ApJ...898...45T} are characterized by high gas and dust content, fueling vigorous star formation activity \citep{2023ApJ...952...76M}. Radiation from massive, young stars in these systems ionizes the surrounding gas, producing prominent emission features that remain observable out to high redshifts. EELGs exhibit not only high [O III] $\lambda$ 5007 line intensities but also large equivalent widths (EWs), typically exceeding 100~\AA. \citet{2013ApJ...778L..22M} confirmed the presence of intense, recent starburst episodes in a sample of EELGs, ruling out significant contributions from older stellar populations in systems with rest-frame [O III] 5007 EWs greater than 500~\AA. These results support the interpretation that EELGs are undergoing vigorous, short-lived bursts of star formation.

%Several studies focused on $z > 1$ found an abundant population of low-mass, low-metallicity with extremely high EWs including \citep{2011ApJ...742..111V,2012ApJ...758L..17B,2014ApJ...785..153M,2010ApJ...719.1168E}.

Building on these insights, the direct measurement of oxygen abundance has traditionally served as a cornerstone in understanding galaxy evolution, due to its strong correlation with both stellar mass and star formation rate. Empirical relations—such as the mass–metallicity and fundamental metallicity relations—have provided valuable constraints on the interplay between gas inflows, star formation, and outflows across galaxies of varying morphologies and evolutionary stages \citep[see, e.g.,][and references therein]{2024ApJ...968...98A}. While oxygen remains the most commonly used tracer of gas-phase metallicity, recent studies have increasingly focused on additional elements—such as argon (Ar), sulfur (S), neon (Ne), and nitrogen (N)—to gain a more comprehensive understanding of chemical enrichment and nucleosynthetic processes in galaxies \citep{2022ApJ...940L..23A, 2023ApJ...959..100I, 2023ApJ...951L..17J, 2023ApJS..269...33N, 2024A&A...681A..30M}.

Interestingly, it was recently demonstrated that the Ar/O ratio exhibits behavior analogous to the well-known [$\alpha$/Fe] abundance pattern, offering a new diagnostic for star formation timescales and chemical enrichment histories \citep{2022A&A...666A.109A}. \citet{2024ApJ...962...50W} investigated several elemental abundance ratios in a population of metal-poor systems using Keck/LRIS spectroscopy, and compared them with chemical abundances derived from recent JWST observations of high-redshift galaxies. These comparisons offer valuable constraints on early chemical evolution and the nucleosynthetic origin of various elements across cosmic time.

Together, these studies motivate a broader approach to chemical enrichment, emphasizing the importance of incorporating multiple elements to gain a more nuanced understanding of galactic metallicity evolution. With the advent of high-resolution spectroscopic surveys, we are now in a golden era for directly measuring a wide range of elemental abundance ratios. When combined with detailed information on SFHs, these measurements provide powerful constraints on models of galaxy evolution, enabling us to probe gas inflow and outflow processes, star formation efficiency, and the metallicity of infalling gas.

In this study, we investigate the chemical enrichment of a sample of extreme emission-line galaxies (EELGs) selected from the Dark Energy Spectroscopic Instrument (DESI) Data Release 1 (DR1) survey \citep{2024AJ....167...62D}. We construct a high-quality sample by requiring stellar masses $M_* \geq 10^7 \mathrm{M}_{\odot}$, large equivalent widths in H$\alpha$ and [O III] $\lambda5007$ ($\mathrm{EW} \geq 500~\text{\AA}$), and simultaneous detections of 19 ionic emission lines with signal-to-noise ratios $\mathrm{S/N} \geq 4$, enabling a robust multi-element analysis of O, N, Ne, S, and Ar abundances. For each system, we combine flexible, non-parametric star formation histories derived from SED fitting with a single-zone chemical evolution model \citep{2020MNRAS.498.1364J}, and develop a multi-element Bayesian framework to forward-model the evolution of abundance ratios. Our approach incorporates time-dependent star formation efficiency, outflow mass-loading, and a physically motivated prescription for inflow metallicity that transitions from primordial to recycled gas, allowing us to directly connect the inferred chemical properties to the underlying baryon cycling processes. We find that EELGs are characterized by short depletion timescales and large mass-loading factors, consistent with burst-driven star formation and rapid gas cycling, and that different physical processes leave distinct, interpretable signatures in abundance space. In particular, we show that combining multiple elemental abundance ratios provides a promising diagnostic for disentangling the roles of star formation, outflows, and inflow enrichment in shaping galaxy evolution.

The structure of this paper is as follows. In Section~\ref{sec:desi-observation}, we describe the DESI observations, the DR1 data products used in this work, and our sample selection. In Section~\ref{sec:SFH-DESI}, we outline the methodology used to derive the star-formation histories of the galaxies. Section~\ref{sec:Abundance-ratio} presents the measurements of elemental abundance ratios. In Section~\ref{sec:GCE-models}, we introduce the single-zone chemical evolution model used to track the evolution of abundances. Section~\ref{Bayesian-galactic-param} describes the Bayesian inference framework and presents the results of the MCMC analysis. Finally, our main conclusions are summarized in Section~\ref{sec:Conclusion}. An optical image gallery is provided in Appendix~\ref{sec:appendix-galleries}. 
%Additional technical details are provided in Appendices~\ref{sec:GCE} and \ref{sec:Corners}.

%%%%%%%%%%%%%%%%%%%%%%%%%%%%%%%%%%%%%%%%%%%%%%%%%%%

\begin{deluxetable*}{r l c c c c c c c}
\tabletypesize{\scriptsize}
\tablecaption{Summary of the DESI EELG sample used in this work. Columns list DESI target identifier, redshift, stellar mass from our \texttt{BAGPIPES} SED fitting reported as median with 16th/84th percentile uncertainties, and gas-phase abundance ratios (1$\sigma$) derived from the emission-line analysis described in Section~\ref{sec:Abundance-ratio}. The initial line-selected candidate list contained one additional object (ID 39628130113032719) that was excluded after visual/SED-quality checks; we do not include it in the final analysis.}
\label{Tab:sample-summary}
\tablehead{
\colhead{N} &
\colhead{DESI ID} &
\colhead{$z$} &
\colhead{$\log_{10}(M_*/M_\odot)$} &
\colhead{$12+\log_{10}(\mathrm{O/H})$} &
\colhead{$\log_{10}(\mathrm{Ar/O})$} &
\colhead{$\log_{10}(\mathrm{N/O})$} &
\colhead{$\log_{10}(\mathrm{Ne/O})$} &
\colhead{$\log_{10}(\mathrm{S/O})$}
}
\startdata
1  & 39627696782709779 & 0.0302 & $7.461^{+0.334}_{-0.287}$ & 8.06$\pm$0.07 & -2.44$\pm$0.11 & -1.46$\pm$0.13 & -0.68$\pm$0.10 & -1.73$\pm$0.10 \\
2  & 39627754429221186 & 0.0233 & $7.815^{+0.179}_{-0.246}$ & 7.97$\pm$0.08 & -2.44$\pm$0.14 & -1.49$\pm$0.15 & -0.66$\pm$0.11 & -1.70$\pm$0.12 \\
3  & 39627758254424582 & 0.0274 & $8.357^{+0.228}_{-0.257}$ & 7.99$\pm$0.05 & -2.24$\pm$0.21 & -1.19$\pm$0.17 & -0.68$\pm$0.07 & -1.56$\pm$0.22 \\
4  & 39627781495065921 & 0.0128 & $8.032^{+0.174}_{-0.174}$ & 7.89$\pm$0.05 & -2.16$\pm$0.11 & -1.33$\pm$0.11 & -0.69$\pm$0.07 & -1.41$\pm$0.10 \\
5  & 39627800457515303 & 0.0134 & $8.906^{+0.097}_{-0.176}$ & 7.93$\pm$0.06 & -2.32$\pm$0.14 & -1.40$\pm$0.14 & -0.68$\pm$0.09 & -1.64$\pm$0.14 \\
6  & 39627809131339694 & 0.0179 & $9.076^{+0.186}_{-0.191}$ & 7.86$\pm$0.06 & -2.25$\pm$0.14 & -1.33$\pm$0.14 & -0.68$\pm$0.09 & -1.56$\pm$0.13 \\
7  & 39627820229463249 & 0.0236 & $7.951^{+0.139}_{-0.172}$ & 7.90$\pm$0.05 & -2.31$\pm$0.11 & -1.37$\pm$0.13 & -0.65$\pm$0.07 & -1.65$\pm$0.11 \\
8  & 39627845927964294 & 0.0202 & $8.533^{+0.213}_{-0.224}$ & 7.91$\pm$0.05 & -2.26$\pm$0.11 & -1.31$\pm$0.12 & -0.62$\pm$0.07 & -1.60$\pm$0.11 \\
9  & 39627851615437621 & 0.0192 & $7.135^{+0.258}_{-0.205}$ & 8.03$\pm$0.07 & -2.22$\pm$0.14 & -1.37$\pm$0.14 & -0.64$\pm$0.11 & -1.47$\pm$0.14 \\
10 & 39627857739124014 & 0.0209 & $8.253^{+0.203}_{-0.252}$ & 8.09$\pm$0.06 & -2.25$\pm$0.14 & -1.35$\pm$0.14 & -0.64$\pm$0.07 & -1.59$\pm$0.14 \\
11 & 39627881336275553 & 0.0291 & $8.018^{+0.207}_{-0.241}$ & 7.99$\pm$0.07 & -2.26$\pm$0.14 & -1.30$\pm$0.13 & -0.61$\pm$0.10 & -1.57$\pm$0.12 \\
12 & 39627933463086197 & 0.0262 & $7.549^{+0.226}_{-0.310}$ & 8.01$\pm$0.10 & -2.27$\pm$0.18 & -1.24$\pm$0.16 & -0.68$\pm$0.16 & -1.54$\pm$0.17 \\
13 & 39627936269076611 & 0.0261 & $7.585^{+0.249}_{-0.301}$ & 8.01$\pm$0.10 & -2.50$\pm$0.19 & -1.46$\pm$0.17 & -0.66$\pm$0.16 & -1.86$\pm$0.17 \\
14 & 39627952844964629 & 0.0286 & $8.108^{+0.193}_{-0.262}$ & 8.08$\pm$0.12 & -2.34$\pm$0.17 & -1.47$\pm$0.20 & -0.71$\pm$0.20 & -1.62$\pm$0.16 \\
15 & 39628069920572605 & 0.0207 & $8.648^{+0.172}_{-0.140}$ & 8.20$\pm$0.06 & -2.32$\pm$0.16 & -1.26$\pm$0.14 & -0.61$\pm$0.08 & -1.61$\pm$0.19 \\
16 & 39628159355717414 & 0.0239 & $7.710^{+0.143}_{-0.180}$ & 7.84$\pm$0.07 & -2.27$\pm$0.13 & -1.40$\pm$0.14 & -0.64$\pm$0.11 & -1.58$\pm$0.11 \\
17 & 39628272547399575 & 0.0252 & $7.280^{+0.353}_{-0.419}$ & 8.04$\pm$0.07 & -1.93$\pm$0.15 & -1.14$\pm$0.12 & -0.71$\pm$0.11 & -1.20$\pm$0.14 \\
18 & 39628506996409541 & 0.0233 & $6.867^{+0.068}_{-0.028}$ & 8.00$\pm$0.10 & -2.10$\pm$0.18 & -1.25$\pm$0.18 & -0.66$\pm$0.16 & -1.42$\pm$0.17 \\
19 & 39633345130267196 & 0.0183 & $7.664^{+0.146}_{-0.197}$ & 7.80$\pm$0.05 & -2.26$\pm$0.10 & -1.22$\pm$0.11 & -0.69$\pm$0.06 & -1.52$\pm$0.11 \\
20 & 39633358333936033 & 0.0267 & $6.984^{+0.185}_{-0.112}$ & 8.24$\pm$0.12 & -2.39$\pm$0.15 & -1.40$\pm$0.16 & -0.60$\pm$0.19 & -1.66$\pm$0.15 \\
21 & 39633358820476651 & 0.0173 & $7.587^{+0.247}_{-0.285}$ & 7.77$\pm$0.06 & -2.22$\pm$0.19 & -1.20$\pm$0.17 & -0.70$\pm$0.08 & -1.59$\pm$0.19 \\
22 & 39633362951868711 & 0.0123 & $8.209^{+0.174}_{-0.174}$ & 8.02$\pm$0.06 & -2.23$\pm$0.15 & -1.32$\pm$0.15 & -0.66$\pm$0.09 & -1.58$\pm$0.15 \\
23 & 39633374372955894 & 0.0230 & $8.074^{+0.210}_{-0.261}$ & 8.12$\pm$0.08 & -2.33$\pm$0.15 & -1.31$\pm$0.15 & -0.63$\pm$0.11 & -1.59$\pm$0.16 \\
\enddata
\end{deluxetable*}

%%%%%%%%%%%%%%%%%%%%%%%%%%%%%%%%%%%%%%%%%%%%%%%%%%%%
\section{DESI observations}
\label{sec:desi-observation}
We use spectroscopy from the Dark Energy Spectroscopic Instrument (DESI; \citealt{2022AJ....164..207D,2024AJ....167...62D}) Data Release 1 (DR1; \citealt{2025arXiv250314745D}). For target selection and baseline measurements we adopt the DR1 value-added catalogs (VACs; see the DR1 release documentation\footnote{\url{https://data.desi.lbl.gov/doc/releases/dr1/}}), in particular the Stellar Mass and Emission Line catalog. In this VAC, stellar masses are estimated with CIGALE using $g/r/z$ photometry from the DESI Legacy Imaging Surveys, WISE W1/W2, and spectrophotometric pseudo-bands constructed from the DESI spectra, while emission-line fluxes are measured via Gaussian fitting with stellar absorption corrections from continuum modeling (see \citealt{2024ApJ...961..173Z} and references therein). We use these VAC products to define the EELG sample and to initialize the subsequent SED fitting and chemical-enrichment analysis.

Our sample selection is based on several well-defined criteria, as outlined below. 
First, to ensure accurate abundance measurements for our planned elements (O, Ar, N, Ne, and S), we require that the line fluxes for all ions listed in Table~\ref{Tab:ion-list1} must have signal-to-noise ratios, S/N $\geq$ 4, where the choice of the auroral lines relies on their capabilities on determining the gas temperature while others are used to estimate the abundance ratios. Second, to reduce the mass diversity while maintaining relevance to our scientific goals, we impose a lower stellar mass limit of $M_* \geq 10^7 M_{\odot}$ to make sure that the system can be described by single zone models. Third, to identify the EELGs galaxies we require:
\begin{align}
\mathrm{EW(H \alpha)}  \geq 500 \mathring{A},  \nonumber\\
\mathrm{EW([O III] ~ 5007)}  \geq  500 \mathring{A}.
\end{align}
This completes our sample selection criteria. Applying these criteria yields a final sample of 23 relatively massive, intensely star-forming dwarf galaxies summarized in Table~\ref{Tab:sample-summary}.
The optical spectral gallery for the DESI EELG sample is shown as Figure~\ref{fig:Galaxy_Spectrum_DESI}, and the optical image gallery is show in Appendix~\ref{sec:appendix-galleries}.

\begin{figure*}[th!]
\center
\includegraphics[width=1.0\textwidth]{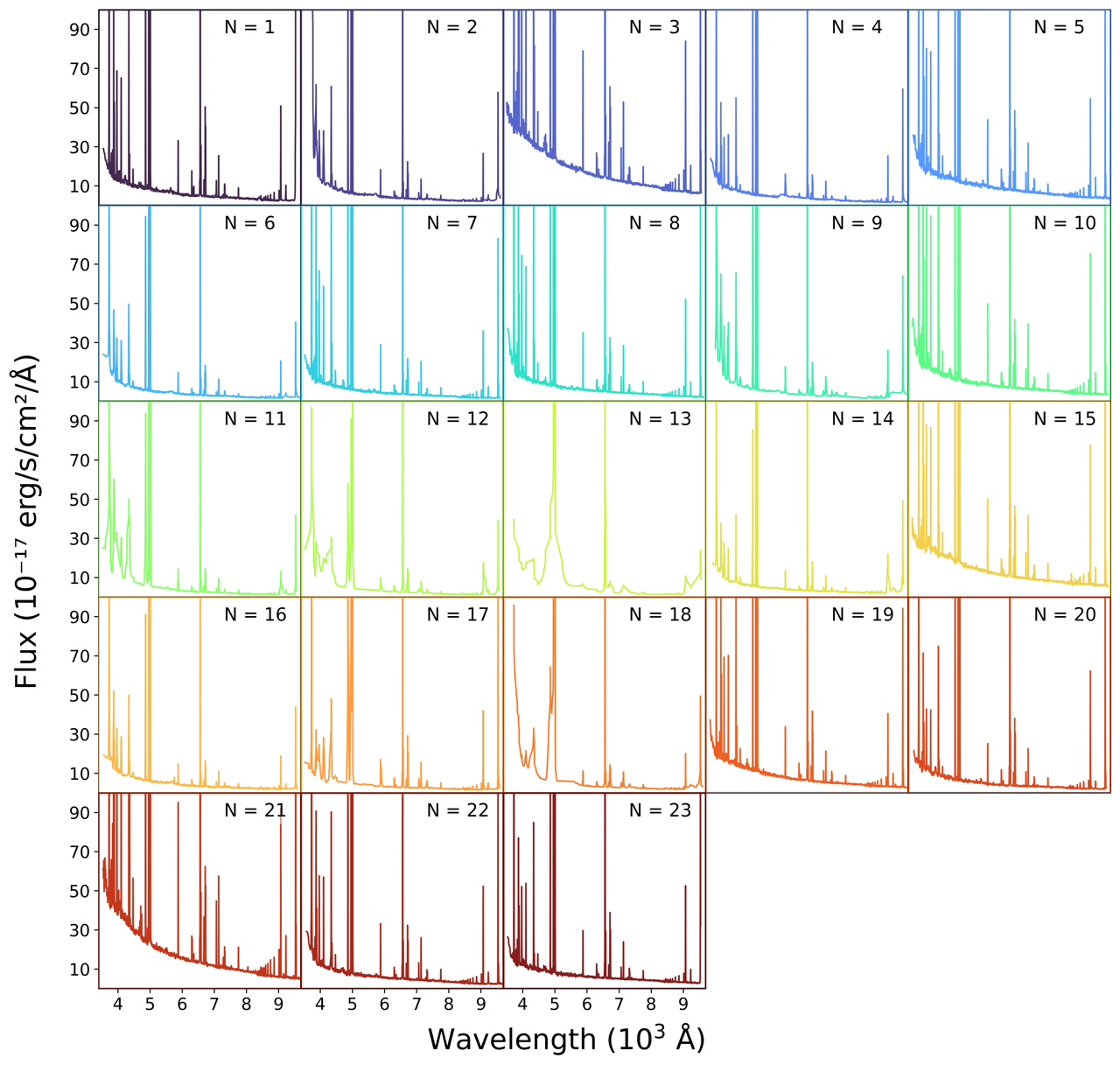}
\caption{Gallery of our EELG spectra in our DESI sample with stellar masses satisfying $M_* \geq 10^{7}M_{\odot}$. }  
\label{fig:Galaxy_Spectrum_DESI}
\end{figure*}

\section{Reconstructing the Star-Formation Histories of Our DESI Galaxy Sample}
\label{sec:SFH-DESI}
We require star-formation histories (SFHs) for each EELG to forward-model chemical enrichment with our single-zone GCE framework \citep{2020MNRAS.498.1364J}. The DR1 VAC SED fits provide useful baseline masses, but their low-dimensional SFH parameterizations are not designed to capture the recent, rapidly varying bursts expected in EELGs. We therefore re-fit the SEDs with a flexible, non-parametric SFH model to obtain SFH realizations and stellar-mass posteriors used throughout the rest of this paper (Table~\ref{Tab:sample-summary}).

\subsection{SED fitting by \texttt{BAGPIPES}}

\texttt{BAGPIPES} \citep{Carnall2018,Carnall2019} provides a Bayesian SED-fitting framework that self-consistently combines stellar population synthesis, nebular emission, dust attenuation, and dust emission. We adopt a non-parametric SFH with a continuity prior \citep{Carnall2019} to capture rapid, bursty variability without overfitting at early times. We fit DESI Legacy Imaging $g/r/z$ photometry, WISE W1/W2, and ten spectrophotometric pseudo-bands constructed from the DESI spectra (following the DR1 VAC approach; \citealt{DESIDR1,Boquien2019}).

The adopted \texttt{BAGPIPES} configuration (data vectors, model components, and prior ranges) is summarized in Table~\ref{Tab:bagpipes-config}.

\begin{table*}[t]
\centering
\caption{\texttt{BAGPIPES} configuration adopted for the DESI EELG sample.}
\label{Tab:bagpipes-config}
\scriptsize
\begin{tabular}{l l}
\hline
\hline
Component & Adopted setup / prior ranges \\
\hline

Photometry &
\parbox[t]{0.75\textwidth}{
DESI Legacy Imaging $g,r,z$ and WISE W1/W2 (all objects).
} \\

Spectroscopy &
\parbox[t]{0.75\textwidth}{
Ten spectrophotometric pseudo-bands integrated from DESI spectra (hat-car windows; uncertainties scaled to be commensurate with broadband photometry; \citealt{DESIDR1,Boquien2019}).
} \\

Dust attenuation &
\parbox[t]{0.75\textwidth}{
Two-component CF00 model \citep{CharlotFall2000}: $A_V\in[0,1]$ mag, $n\in[0.1,1]$, $\mu\in[0.2,0.5]$, $\eta\in[1.5,3.0]$.
} \\

Dust emission &
\parbox[t]{0.75\textwidth}{
DL07 templates \citep{DraineLi2007}: $q_{\rm PAH}\in[0.5,4.7]$, $U_{\rm min}\in[0.1,25]$, $\alpha\in[1.5,3.0]$, $\gamma\in[0,0.2]$.
} \\

Nebular emission &
\parbox[t]{0.75\textwidth}{
$\log_{10}U\in[-3.5,-1]$, $n_e\in[50,500]~\mathrm{cm}^{-3}$, $Z_{\rm gas}\in[10^{-4},10^{-2}]$, $f_{\rm esc}=0$.
} \\

SFH &
\parbox[t]{0.75\textwidth}{
Non-parametric SFH with continuity prior \citep{Carnall2019}; lookback-time bin edges (Myr): $[0.1,0.3,1,5,20,100,500,1000,3000,10^{4},12000]$; $\log_{10}(M_*/M_\odot)\in[4,10]$; $Z_*/Z_\odot\in[10^{-3},3]$.
} \\

Additional &
\parbox[t]{0.75\textwidth}{
Birth-cloud timescale $t_{\rm BC}\in[0.007,0.02]$ Gyr; velocity dispersion ${\rm veldisp}\in[10,300]~\mathrm{km\,s^{-1}}$.
} \\

Noise model &
\parbox[t]{0.75\textwidth}{
Systematic floors added in quadrature: 5\% ($g,r,z$), 10\% (W1,W2), 7\% (continuum pseudo-bands), 15\% (emission-line pseudo-bands); cap $\mathrm{S/N}$ at 20.
} \\

\hline
\end{tabular}
\end{table*}

%%%%%%%%%%%%%%%%%%%%%%%%%%%%%%%%%%%%%%%%%%%%%%%%%%%%%%%%%%%
\paragraph{Noise model and systematic uncertainties:}
We adopt the wavelength-dependent systematic error model listed in Table~\ref{Tab:bagpipes-config}, added in quadrature to the reported measurement uncertainties, and cap the maximum $\mathrm{S/N}$ at 20 to prevent unrealistically small errors from dominating the likelihood.

%We jointly fit DESI $g$, $r$, and $z$ broadband photometry, WISE W1 and W2 photometry (present for all objects in our sample), and ten pseudo–narrow-band spectrophotometric measurements derived from the DESI spectra using a hat-car (hat-carriage) pseudo-filter approach. 

Systematic floors and the $\mathrm{S/N}$ cap are also listed in Table~\ref{Tab:bagpipes-config}.

Appendix~\ref{sec:appendix-galleries} depicts the optical image gallery for our sample.  In each panel, the yellow circle indicates the region of the galaxy targeted by DESI for spectroscopic observations. It is noted that the spectroscopy and photometry trace different parts of the galaxy. 

A notable diversity in both of the morphology and environment is seen in our EELG DESI sample, where some dwarf galaxies are embedded within more massive halos, whereas others appear to be nearly isolated systems.

%%%%%%%%%%%%%%%%%%%%%%%%%%%%%%%%%%%%%%%%%%%%%%%%%%%%
\begin{figure*}[th!]
\center
\includegraphics[width=1.0\textwidth]{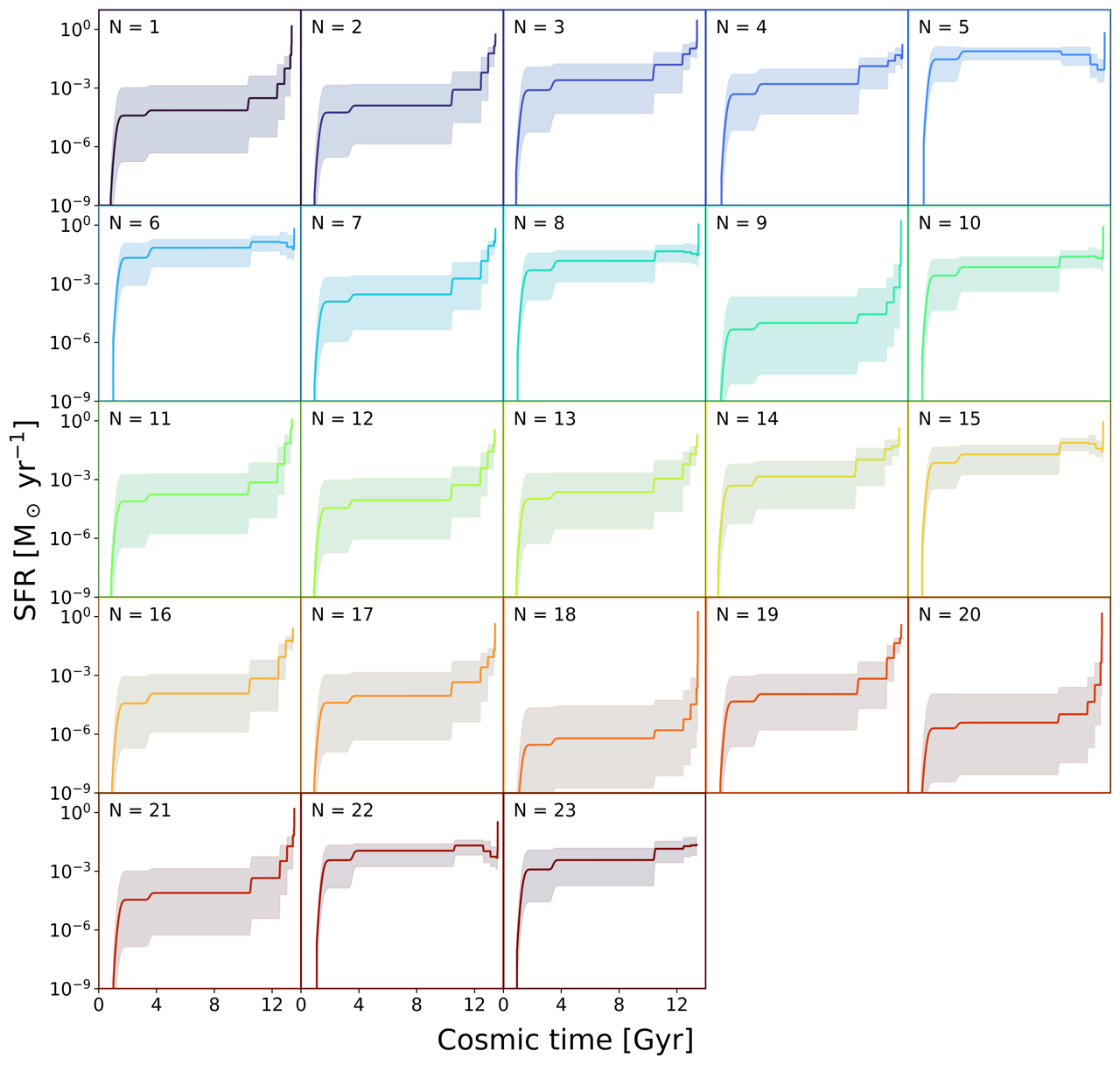}
\caption{Inferred star formation histories (SFHs) for the DESI EELG sample derived from our \texttt{BAGPIPES} SED fitting. In each panel, the solid curve shows the median (50th percentile) SFH, while the shaded region denotes the 16th–84th percentile credible interval from the non-parametric posterior distribution. All galaxies display relatively modest star formation at early times followed by a pronounced recent burst toward the epoch of observation, consistent with the burst-dominated evolutionary phase expected for EELGs.}  
\label{Fig:SFH-fit}
\end{figure*}
%%%%%%%%%%%%%%%%%%%%%%%%%%%%%%%%%%%%%%%%%%%%%%%%%%%%%

Figure \ref{fig:Galaxy_Spectrum_DESI} presents the gallery of galaxy spectra in our DESI sample. 

%%%%%%%%%%%%%%%%%%%%%%%%%%%%%%%%%%%%%%%%%%
\subsection{Inferred SFH from SED fitting}
Having described our SED fitting methodology, we now present the resulting inferred physical properties. Table~\ref{Tab:sample-summary} lists the galaxy identifiers, redshifts, and stellar-mass posteriors (16th/50th/84th percentiles) inferred from our \texttt{BAGPIPES} fits.

As noted above, one galaxy (ID = 39628130113032719) was excluded from the final sample. Its SED fit yielded a statistically poor solution, and the inferred star formation history did not exhibit the bursty behavior characteristic of the rest of the sample. Moreover, this galaxy is spatially and spectroscopically close to another object in the sample, raising the possibility of contamination or other spectroscopic systematics. Although a detailed investigation of these effects would be worthwhile, it lies beyond the scope of the present study. We therefore adopt a conservative approach and exclude this galaxy from the analysis.

Figure~\ref{Fig:SFH-fit} presents the gallery of inferred star formation histories obtained from our \texttt{BAGPIPES} SED fitting. In each panel, we show the 16th, 50th, and 84th percentiles of the posterior SFH derived from the non-parametric model. In all cases, the galaxies exhibit relatively low and nearly constant star formation at early times, followed by a pronounced recent burst toward the epoch of observation. This behavior is consistent with our expectation that EELGs undergo late-time bursty star formation. As we demonstrate in the following sections, this recent burst plays a central role in shaping the subsequent metallicity enrichment analysis.

\section{Element Abundance ratio}
\label{sec:Abundance-ratio}
To calculate the total abundance of an element 
$X$ in an ionized nebula, we first determine the electron temperature and electron number-density by analyzing the appropriate emission lines in the gas phase. These physical conditions are then used, in combination with the relevant emission lines, to compute the abundances of different ionization states for each element. Below, we outline the methodology used to estimate the abundance ratios for several elements.

\subsection{Estimation of the abundance ratios}
\label{Abundance-Ratio-Calculation}
The abundance ratio for element 
$X$ is estimated as:
\begin{equation}
\label{ICF_formula}
\frac{X}{H} = \mathrm{ICF}(X^{+i}) \frac{X^{+i}}{H^+}.
\end{equation}
where $X^{+i}$ refers to the specific ions whose ionic abundance can be directly determined from the observed emission lines, while the ionization correction factor (ICF) accounts for the contribution of unobserved ionization states \citep{1969BOTT....5....3P,1977MNRAS.179..217P,1978A&A....66..257S,1985ApJ...291..247M,2006A&A...448..955I,2007MNRAS.381..125P,2013MNRAS.432.2512D,2021MNRAS.505.2361A}. 

Among various ionization correction factor (ICF) prescriptions that are available in the literature, \citet{2025A&A...697A..61E} evaluated several of them by computing the abundance ratios of multiple elements using different ICFs. After a detailed comparison of the resulting values, they concluded that the methodology of \citet{2006A&A...448..955I} provides a more consistent and reliable result across a wide range of abundance ratios. Accordingly, in this study, we adopt the standard ICF prescription proposed by \citet{2006A&A...448..955I}:
%%%%%%%%%%%%%%%%%%%%%%%%%%%%%%%%%%%%
\begin{equation}
\label{ICF_Ar}
\mathrm{ICF(Ar^{+2})} = 
\begin{cases}
0.278 v + 0.836 + 0.051/v,  & \mathrm{low}\ Z \\
0.285 v+ 0.833 + 0.051/v, & \mathrm{interm}\ Z \\
0.517 v + 0.763 + 0.042/v, & \mathrm{high}\ Z  \\
\end{cases}
\end{equation}
%%%%%%%%%%%%%%%%%%%%%%%%%%%%%%%%%%%%
\begin{equation}
\label{ICF_N}
\mathrm{ICF(N^{+})} = 
\begin{cases}
-0.825 v + 0.718 + 0.853/v,  & \mathrm{low}\ Z \\
-0.809 v+ 0.712 + 0.852/v, & \mathrm{interm}\ Z \\
-1.476 v + 1.752 + 0.688/v, & \mathrm{high}\ Z
\end{cases}
\end{equation}
%%%%%%%%%%%%%%%%%%%%%%%%%%%%%%%%%%%%
\begin{equation}
\label{ICF_S}
\mathrm{ICF(S^{+} + S^{+2})} = 
\begin{cases}
0.121 v + 0.511 + 0.161/v,  & \mathrm{low}\ Z \\
0.155 v+ 0.849 + 0.062/v, & \mathrm{interm}\ Z \\
0.178 v + 0.610 + 0.153/v, & \mathrm{high}\ Z
\end{cases}
\end{equation}
%%%%%%%%%%%%%%%%%%%%%%%%%%%%%%%%%%%%
\begin{equation}
\label{ICF_Ne}
\mathrm{ICF(Ne^{+2})} = 
\begin{cases}
-0.385 w + 1.365 + 0.022/w,  & \mathrm{low}\ Z \\
-0.405 w+ 1.382 + 0.021/w, & \mathrm{interm}\ Z \\
-0.591 w + 0.927 + 0.546/w, & \mathrm{high}\ Z 
\end{cases}
\end{equation}
%%%%%%%%%%%%%%%%%%%%%%%%%%%%%%%%%%%%
where we have defined $v$ and $w$ as:
\begin{align}
\label{v-def}
v \equiv \left( \frac{O^{+}}{O^+ + O^{+2}} \right) ~~~~~,~~~~~
w \equiv \left( \frac{O^{+2}}{O^+ + O^{+2}} \right)
\end{align}

To compute the abundance ratio for $O/H$ we use the following formula:  % Make sure to use Mathrm consistently throughout the analysis. 
\begin{align}
\label{OxygenToH}
& \frac{O}{H} = \left( \frac{O^+ + O^{+2}}{H^+} \right),
\end{align}
where we use the list of ions as specified in Table \ref{Tab:ion-list1}. The resulting abundance ratios for all galaxies are summarized in Table~\ref{Tab:sample-summary}.

In computing $\frac{O^+}{H^+}$, $\frac{O^{+2}}{H^+}$, $\frac{Ar^{+2}}{H^+}$, $\frac{S^{+1}}{H^+}$, and $\frac{S^{+2}}{H^+}$ we read their emissivities from the Pyneb code. Incorporating the above ICFs inside Equation  \ref{ICF_formula} we determine the abundance ratios of $Ar/H$, $N/H$, $Ne/H$, and $S/H$.   

For calculating different element abundance ratios we would also need to specify the electron temperature and electron number density of different ions. We estimate them as the following: 
\begin{align}
\label{Line_T}
& \mathrm{T_e}([N II])  = \mathrm{T_e}([O II]), \nonumber\\
& \mathrm{T_e}([S II])  = \mathrm{T_e}([O II]), \nonumber\\
& \mathrm{T_e}([Ne III])  = \mathrm{T_e}([O III]), \nonumber\\
& \mathrm{T_e}([Ar III])  = \mathrm{T_e}([S III]). 
\end{align}
We note that the electron temperature $\mathrm{T}_\mathrm{e}([\mathrm{O III}])$ is determined using the ratio of the nebular lines [O III] 4959 and [O III] 5007 to the auroral line [O III] 4363. Because the electron temperature depends on the electron number density, we begin with an initial estimate of $n_\mathrm{e} = 200~\mathrm{cm}^{-3}$ and iteratively refine this value. At each step, the inferred temperature is used to update the electron density, which is derived from the [O II] doublet ([O II] 3726 and [O II] 3729). Similarly, the electron temperature $\mathrm{T}_\mathrm{e}([\mathrm{O II}])$ is calculated using the ratio of [O II] 3726 to the auroral lines [O II] 7320 and  [O II] 7331.

Ultimately to compute $\mathrm{T_e}$([S III]) we use the inferred electron number density from the calculation above and using a combination of ([S III] 9071, [S III] 9533) and [S III] 6312 lines, respectively. 

To compute the de-reddening factor from the dust attenuation we use the Pyneb package and correct each line using the Balmer decrement, H$\alpha$/H$\beta$, ratio. Each line is then corrected based on its wavelength. 

Finally, to propagate measurement uncertainties into abundance uncertainties, we use an MCMC approach in which each line flux is resampled from a Gaussian centered on the measured value with width set by the flux error. We report the median and 16th/84th percentiles for each abundance ratio; the final values are listed in Table~\ref{Tab:sample-summary}.

%%%%%%%%%%%%%%%%%%%%%%%%%%%%%%%%%%%%%%%%%%%%%%%%%%%%
\begin{figure*}[th!]
\center
\includegraphics[width=0.97\textwidth]{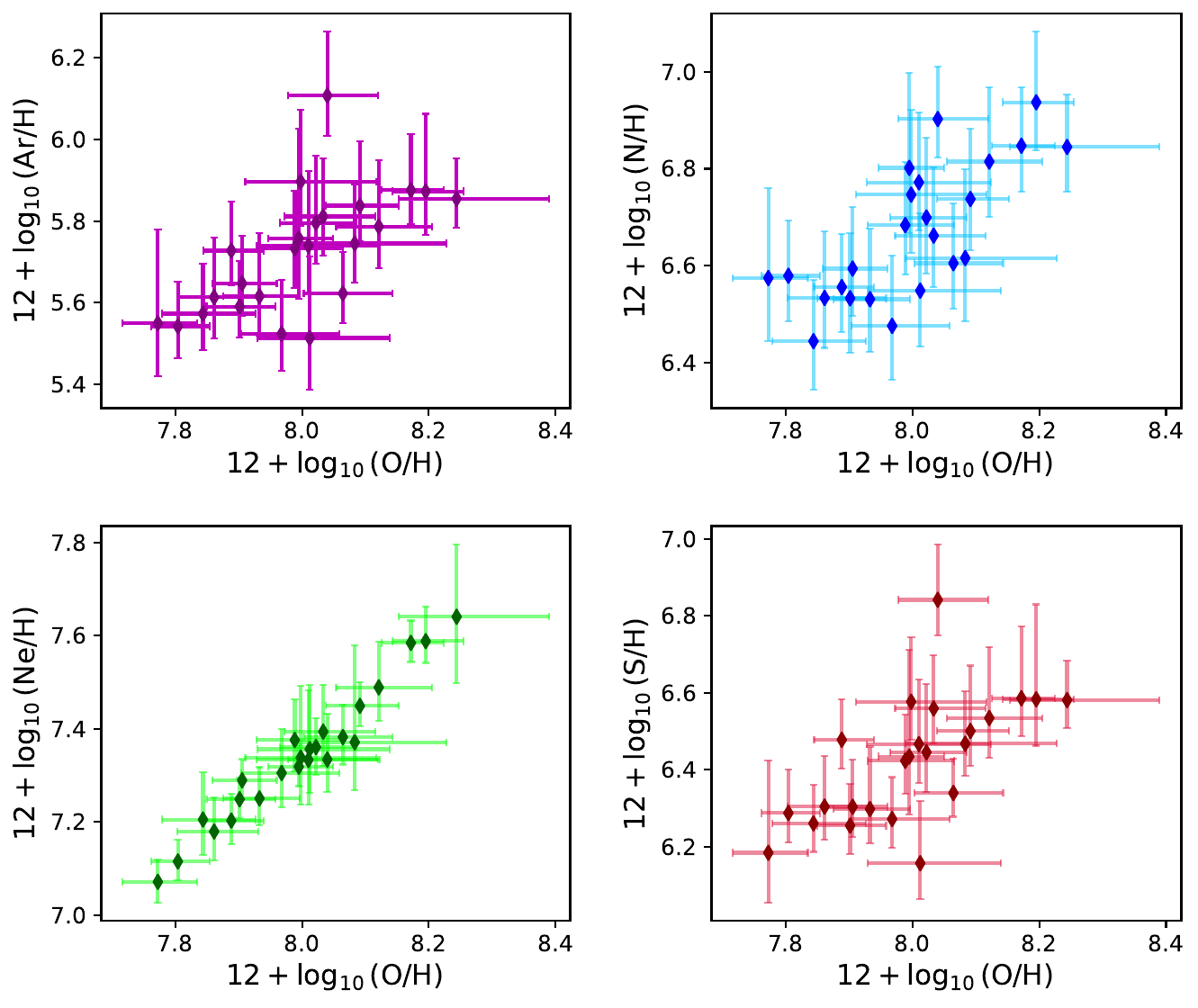}
\caption{Abundance ratios for the Extremely emission dominated galaxies with $M_{*} \geq 10^7 M_{\odot}$  satisfying the criteria in Table \ref{Tab:ion-list1} with S/N $\geq$ 4 (dotted blue). 
}  
\label{fig:Metallicity-sample}
\end{figure*}
%%%%%%%%%%%%%%%%%%%%%%%%%%%%%%%%%%%%%%%%%%%%%%%%%%%%%

\subsection{EW vs metallicity}
\begin{figure}[th!]
\center
\includegraphics[width=0.5\textwidth]{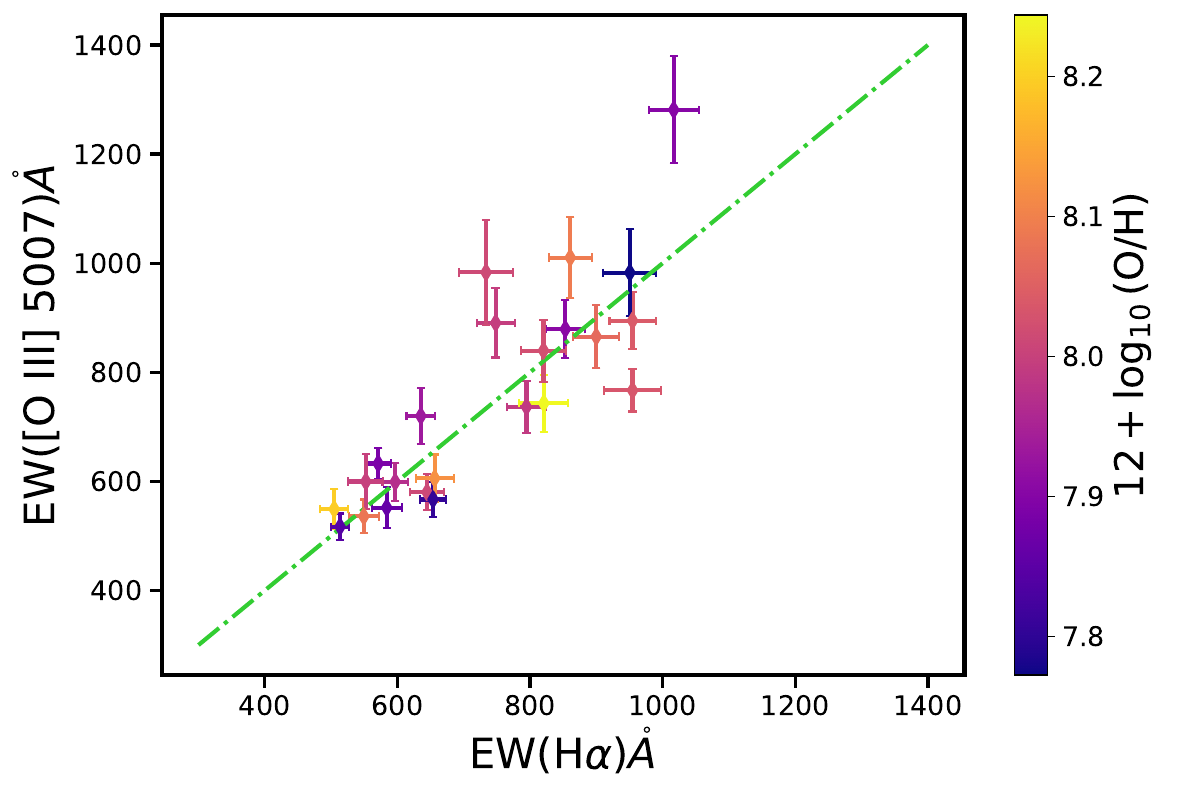}
\caption{The rest frame EWs for EW(H$\alpha$) vs EW([O III] 5007). Color-coded we also present 12 + $\log_{10}{\mathrm{(O/H)}}$ for all of galaxies in our sample.}  
\label{fig:Equivalent-width-metal}
\end{figure}

Figure \ref{fig:Equivalent-width-metal} presents EW(H$\alpha$)  vs.  EW([O III] 5007) in our galaxy sample. 
From the plot it is inferred that EWs are broadly similar between these two lines, where the background 
dotted-dashed line shows equality. 
Furthermore, it is seen that a good fraction of galaxies in our sample have EWs well above the criterion of 500~\AA. As noted in the introduction, such galaxies are typically expected to undergo intense, short-lived bursts of star formation. We will check this explicitly in the next section. The plot is color-coded by the gas phase metallicity as traced by 12 + $\log_{10}{\mathrm{(O/H)}}$. It is clearly seen that there are some levels of gradient for high metallicity regions corresponding to lower EWs, while lower-metallicities show higher EWs. 

Having chosen our galaxy sample, in the following we adopt a customized SED fitting to unravel the SFHs of our sample galaxies, as is relevant to our subsequent analysis of their metallicity enrichment. 

%%%%%%%%%%%%%%%%%%%%%%%%%%%%%%%%%%%%%%%%%%%%%%%%%%%%
%%%%%%%%%%%%%%%%%%%%%%%%%%%%%%%%%%%%%%%%%%%%%%%%%%%%%

\subsection{Mass Metallicity relation}
\label{mass-metallicity-sample}
\begin{figure}[th!]
\center
\includegraphics[width=0.47\textwidth]{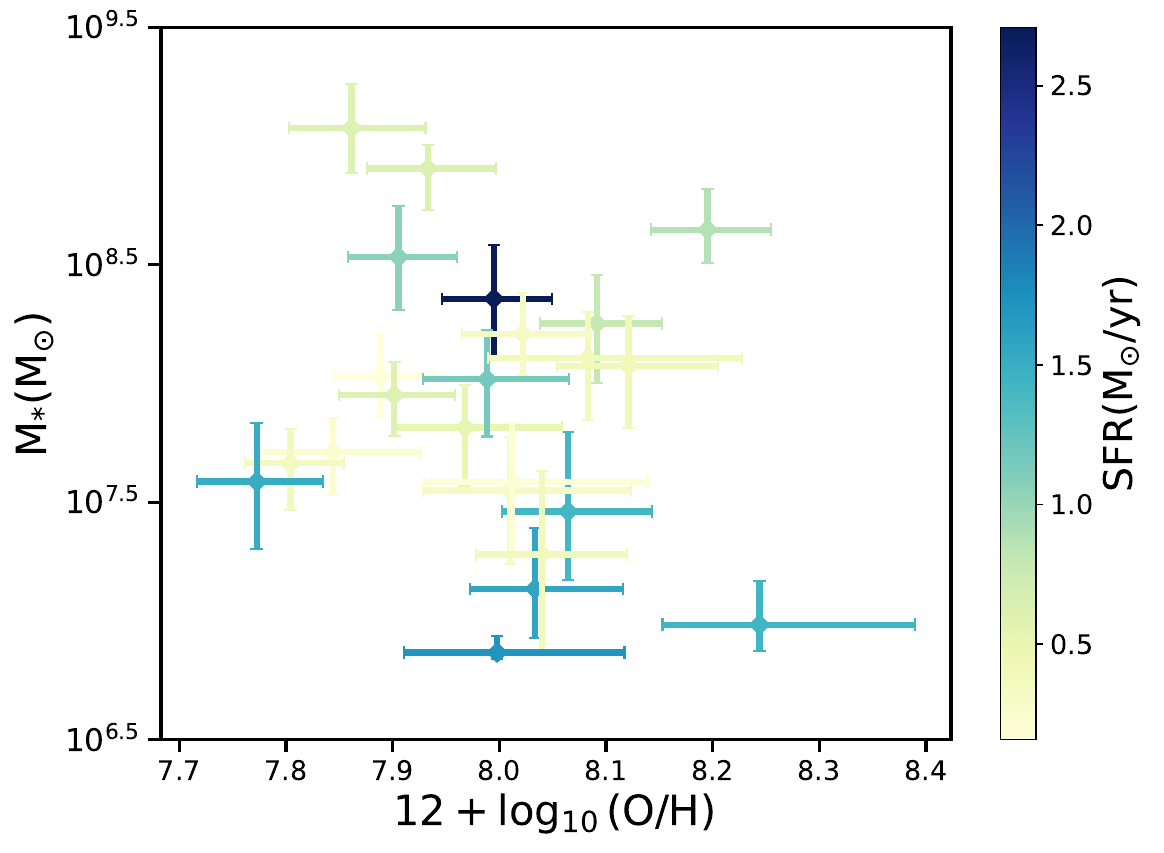}
\caption{Stellar mass ($M_\star$) as a function of gas-phase metallicity ($12 + \log_{10}(\mathrm{O/H})$) for the DESI EELG sample. Points are color-coded by star formation rate (SFR). The galaxies preferentially lie in the low-metallicity, high-SFR region of parameter space, in agreement with the expected properties of extreme emission-line galaxies.}  
\label{fig:Stellar_Mass_Metal_SFR}
\end{figure}

Figure \ref{fig:Stellar_Mass_Metal_SFR} presents the relationship between stellar mass ($M_*$) and gas-phase metallicity, traced by 12 + $\log_{10}{(\mathrm{O/H})}$, for our EELG DESI sample. Each point is color-coded by the corresponding star formation rate (SFR) at the time of observation, expressed in units of $M_\odot/\mathrm{yr}$. The sample displays a narrow range of low metallicities. This outcome is expected given our selection criteria, which favor emission-line dominated galaxies typically characterized by active star formation and chemically enriched interstellar media.

%%%%%%%%%%%%%%%%%%%%%%%%%%%%%%%%%%%%%%%%%%%%%%%%%%%%
%%%%%%%%%%%%%%%%%%%%%%%%%%%%%%%%%%%%%%%%%%%%%%%%%%%%%

\section{Galactic Chemical Evolution models}
\label{sec:GCE-models}

%% Add more here ... 

Throughout our analysis, we utilize the Versatile Integrator for Chemical Evolution (VICE) code, developed by \citet{2020MNRAS.498.1364J}, to solve the integro-differential equations governing a one-zone chemical evolution model. The time-evolution of the gas-supply is given by \citet{2020MNRAS.498.1364J}: 
\begin{equation}
\label{gas-evolution}
\dot{M}_{\mathrm{g}} =  \dot{M}_{\mathrm{in}} - \dot{M}_* - \dot{M}_{\mathrm{out}} + \dot{M}_{\mathrm{returned}}. 
\end{equation}
Here, $\dot{M}_{\mathrm{in}}$ denotes the mass infall rate, $\dot{M}_*$ represents the star formation rate (SFR), $\dot{M}_{\mathrm{out}}$ corresponds to the mass outflow rate, and $\dot{M}_{\mathrm{returned}}$ describes the rate of stellar mass recycling. Technically, VICE is designed to operate flexibly, allowing the user to specify either the infall history ($\dot{M}_{\mathrm{in}}$), the star formation history ($\dot{M}_*$), or the gas mass history ($M_{\mathrm{g}}$). In addition, it is necessary to define the star formation efficiency timescale, $\tau_* \equiv M_{\mathrm{g}}/\dot{M}_*$, as well as the outflow rate, which is typically parameterized by the dimensionless mass-loading factor, $\eta \equiv \dot{M}_{\mathrm{out}} / \dot{M}_*$. Finally, for stellar mass recycling, VICE adopts a continuous recycling formalism following the standard approach outlined by \citet{2008ApJ...676..594K}, utilizing the initial–final mass relation presented in the same work. Following the standard approach from VICE in our analysis below,  we include the recycling from the core-collapse SNe (CCSNe), the Type Ia SNe, and the AGB stars. (Further details on the implementation can be found in Equation 2 of \citet{2020MNRAS.498.1364J}.)

\subsection{The scaling relation for $\tau_*$ and $\eta$}
\label{tau-eta-scaling}
Since we have already reconstructed the SFHs for all of galaxies in our sample, we adopt the `sfr' mode throughout this work in our single-zone simulations and we utilize our fitted SFHs from Figure \ref{Fig:SFH-fit} as an input to our models. Furthermore, given the bursty nature of star formation in our DESI sample (as illustrated in Figure \ref{Fig:SFH-fit}), we introduce a time-dependent star formation efficiency timescale $\tau_*$ as well as a mass-loading factor $\eta$, each assumed to scale with the SFR as:

\begin{align}
\label{Tau-form}
\tau_*(t) = & ~   A_{\tau}^{(-1/n_{\tau})} \left( \frac{SFR(t)}{M_{\odot}/yr} \right)^{(1-n_{\tau})/n_{\tau}} 
\left(\frac{\mathrm{Myr}}{10^{6}} \right), \\
\label{eta-form}
\eta(t) = & ~ A_{\eta} \left( \frac{SFR(t)}{M_{\odot}/yr} \right)^{n_{\eta}}. 
\end{align}
where in parameterizing $\tau_*$ we have used the Schmidt relation \citep{1959ApJ...129..243S,1963ApJ...137..758S,1998ApJ...498..541K} to connect the rate of the star formation to the gas mass, $\dot{M}_* = A_{\tau} M_{\mathrm{g}}^{n_{\tau}} $, while in the $\eta$ scaling we have assumed that the mass loading factor is tightly connected with the rate of the star formation. 
There are four free parameters in Equations (\ref{Tau-form}-\ref{eta-form}), including $A_{\tau}, n_{\tau}, A_{\eta}$, and $n_{\eta}$. 

\subsection{The scaling relation for $Z_{\mathrm{in}}$}
\label{Zin-scaling}
Next, we parameterize the infall metallicity. Two points are in order. First, since we only track O, N, Ne, S, and Ar in our analysis we take a fraction of $\sim 0.6032$ of the infall metallicity for our elements. This is subsequently split between individual elements using their atomic number and solar metallicities. 

Second, since in our analysis we cover the full galaxy evolution we need to choose an ansatz that naturally connects the primordial gas infall metallicity to a late-time recycled gas. In our analysis we adopt the following expression:
\begin{equation}
\label{z-infall}
Z_{\mathrm{in}}(t) = Z_{\mathrm{prim}} + \left( Z_{\mathrm{rec}} - Z_{\mathrm{prim}} \right) S(t),
\end{equation}
where we adopt a simple parameterized $S(t)$ as:
\begin{equation}
\label{S-function}
S(t) \equiv \frac{1}{1 + \exp{\left(-(t - t_{50})/w \right)}},
\end{equation}
where $t_{50}$ refers to the time where 50\% of stars are formed, which is given by equating the formed stellar mass fraction $F(t)$ to 1/2. 
\begin{equation}
F(t) = \frac{\int_{t_0}^{t} SFR(t) dt }{\int_{t_0}^{t_{obs}} SFR(t) dt},
\end{equation}
Furthermore, w is defined as the time-scale for a switch from 20\% to 80\% of stars and it is given by:
\begin{equation}
w \equiv \frac{t_{80} - t_{20}}{2.77}.
\end{equation}
The above ansatz for the infall metallicity brings two additional parameters into the pool; i.e. $Z_{\mathrm{prim}}$ and  $Z_{\mathrm{rec}}$. We have checked that a small primordial infall metallicity value does not really affect the late-time metallicity enrichment. Consequently we fix $Z_{\mathrm{prim}} = 10^{-5}$ and in our analysis below only explore the role of the recycled metallicity, i.e. $Z_{\mathrm{rec}}$. 

Having presented our full ansatz for the metallicity enrichment, we next explore a five-dimensional parameter space in the MCMC analysis. Our goal is to reproduce the present-day abundance ratios in Table~\ref{Tab:sample-summary} while simultaneously matching their implied enrichment trajectories.

%%%%%%%%%%%%%%%%%%%%%%%%%%%%%%%%%%%%%%%%%%%%%%%%%%%
It was noted by \citet{2020MNRAS.498.1364J} that bursty star formation may occur either in young dwarf galaxies or as a consequence of localized gas accretion within the spiral arms of massive galaxies. Such localized variations in star formation frequently lead to the emergence of loops in various elemental abundance ratio diagrams.

We adopt CCSNe yields from \citet{2018ApJS..237...13L} (evaluated at [M/H] = -1 and 0), Type Ia yields from the W70 model of \citet{1999ApJS..125..439I}, and AGB yields from \citet{Karakas2010}.

\subsection{Different set of yields tables}
\label{sec:yields}
To incorporate the role of CCSNe, Type Ia SNe, and the AGB stars we have used standard yield tables from the literature. Below, we describe our selections for these yields and their values:

\subsubsection{CCSNe yield}
\label{ccsne-yield-function}
Following the approach of \citet{2020MNRAS.498.1364J} we take an instantaneous explosion approximation for the CCSNe, where it is assumed that the yield of a given element X is ejected out simultaneously at all time-steps. Furthermore, we also assume a min(max) mass value of 8(40) M$_{\odot}$ as the mass limits for producing the CCSNe. We then compute the IMF-averaged yields as relevant to the observations. We check that changing the IMF in this range does not significantly alter the net yields. Consequently, we fix the IMF to be Kroupa throughout this work. 
In our analysis we use the most standard yield set that is being widely used in the literature. In particular, we check if \citet{2018ApJS..237...13L} (with different available metallicities) produces a good fit to the final element abundance ratios from the DESI data. In our analysis we adopt two different values of [M/H] = $\log_{10}{\left((Z/Z_{\odot}) \right)} = -1, 0$.

The yield choices above are used throughout the enrichment modeling.

\subsubsection{Type Ia SNe}
\label{type-Ia-yield}
In our analysis, we model the Type Ia supernova yields using the W70 model from \citet{1999ApJS..125..439I}. We adopt an exponential time-delay distribution with a fiducial delay time of $\tau_{\mathrm{del}} = 2.0$ Gyr. To assess the impact of this assumption, we also explore variations in the delay time and find that they do not significantly influence the final values of the element abundance ratios.

\subsubsection{AGB Stars}
\label{AGB-star-yield}
In our analysis below we fix the AGB yield to be that of \cite{Karakas2010}. 

\section{Galactic parameter inference using a forward modeling Bayesian approach}
\label{Bayesian-galactic-param}
We infer the baryon-cycling parameters in our single-zone chemical-evolution model by forward-modeling the observed present-day abundance ratios in each galaxy. Specifically, we constrain the five free parameters in Equations~\ref{Tau-form}--\ref{z-infall}, $\left(n_{\tau}, \log_{10}(A_{\tau}), n_{\eta}, \log_{10}(A_{\eta}), \log_{10}(Z_{\rm rec})\right)$, using the gas-phase measurements of $12+\log_{10}(\mathrm{O/H})$ and the abundance ratios $\log_{10}(\mathrm{Ar/O})$, $\log_{10}(\mathrm{N/O})$, $\log_{10}(\mathrm{Ne/O})$, and $\log_{10}(\mathrm{S/O})$ (Table~\ref{Tab:sample-summary}). Our inference uses Markov Chain Monte Carlo (MCMC) sampling and propagates uncertainties in the reconstructed star-formation histories (SFHs) by repeating the analysis for the 16th/50th/84th-percentile SFH realizations from \texttt{BAGPIPES}.

Our analysis is based on a Bayesian Monte Carlo Markov Chain (MCMC) approach. Below we describe the key aspects of our MCMC pipeline and summarize the resulting outputs.

\subsection{Inference setup and reproducibility}
\label{sec:mcmc-pipeline}
For a given galaxy and assumed yield set (Section~\ref{sec:yields}), we run VICE in ``sfr'' mode with the fixed input SFH and compute the model abundance ratios as a function of time. Our observed data vector for each object is $\mathbf{A}_{\rm obs} \equiv$ ($12+\log_{10}(\mathrm{O/H}), \log_{10}(\mathrm{Ar/O}), \log_{10}(\mathrm{N/O}),$ $\log_{10}(\mathrm{Ne/O}), \log_{10}(\mathrm{S/O})$), and the model provides an analogous $\mathbf{A}_{\rm mod}(t)$ evaluated along the enrichment history. Because these systems are bursty and the enrichment tracks can fluctuate on short timescales, we evaluate the likelihood over a short window preceding the observational epoch rather than at a single final timestep. Specifically, we define the ``most recent'' comparison interval as the final $\Delta t = 10$ Myr of the simulation and compute a weighted $\chi^2$ misfit in this window (with the highest weights assigned to timesteps closest to the observation).

We adopt a Gaussian likelihood in the five-dimensional abundance space, with observational uncertainties from the emission-line analysis augmented by systematic uncertainties associated with the adopted solar abundance scale (added in quadrature). We adopt solar abundances $12+\log_{10}(\mathrm{O/H})_\odot = 8.69$, $12+\log_{10}(\mathrm{N/H})_\odot = 7.83$, $12+\log_{10}(\mathrm{Ne/H})_\odot = 7.93$, $12+\log_{10}(\mathrm{S/H})_\odot = 7.12$, and $12+\log_{10}(\mathrm{Ar/H})_\odot = 6.40$, with corresponding systematic uncertainties of $(0.05, 0.05, 0.10, 0.03, 0.13)$ dex, respectively.

We adopt broad priors on the five free parameters:
\begin{equation}
\begin{aligned}
\log_{10}(A_{\eta}) &\in [-0.3,\; 3.0] \\
n_{\eta}            &\in [-1.0,\; 1.0] \\
n_{\tau}            &\in [0.9,\; 2.0] \\
\log_{10}(A_{\tau}) &\in [-13.0,\; -7.0] \\
\log_{10}(Z_{\mathrm{rec}}) &\in [-5.0,\; -1.0]
\end{aligned}
\label{priors}
\end{equation}
In addition to these broad bounds, we include soft Gaussian priors to regularize the implied present-day values of the mass-loading factor $\eta$ and depletion time $\tau_\star$ at two SFH-defined reference SFRs (a recent SFR averaged over the final $\sim 30$ Myr and the 70th percentile of the SFR distribution over the full SFH). For $\eta$ we adopt a prior centered on $\log_{10}(\eta)=1.0$ with dispersions of $\sigma=1.0$ and $2.0$ at the two reference SFRs, and for $\tau_\star$ we adopt a prior centered on $\log_{10}(\tau_\star)=0.0$ with dispersions of $\sigma=1.5$ and $2.0$. These priors gently favor physically plausible gas-cycling regimes while remaining weak compared to the information provided by the multi-element abundances.

We sample the posterior with an ensemble sampler ($64$ walkers, $3500$ steps each). Walkers are initialized uniformly within the prior bounds with a small numerical buffer to avoid starting exactly on boundaries. To propagate uncertainties in the SFH, we repeat the inference for the 16th/50th/84th percentile SFH realizations from \texttt{BAGPIPES}. For all summary plots, we combine posterior samples from the three SFH realizations with equal weights.

%%%%%%%%%%%%%%%%%%%%%%%%%%%%%%%%%%%%%%%%%%%%%%%%%%%
\begin{figure*}[th!]
\center
\includegraphics[width=1.0\textwidth]{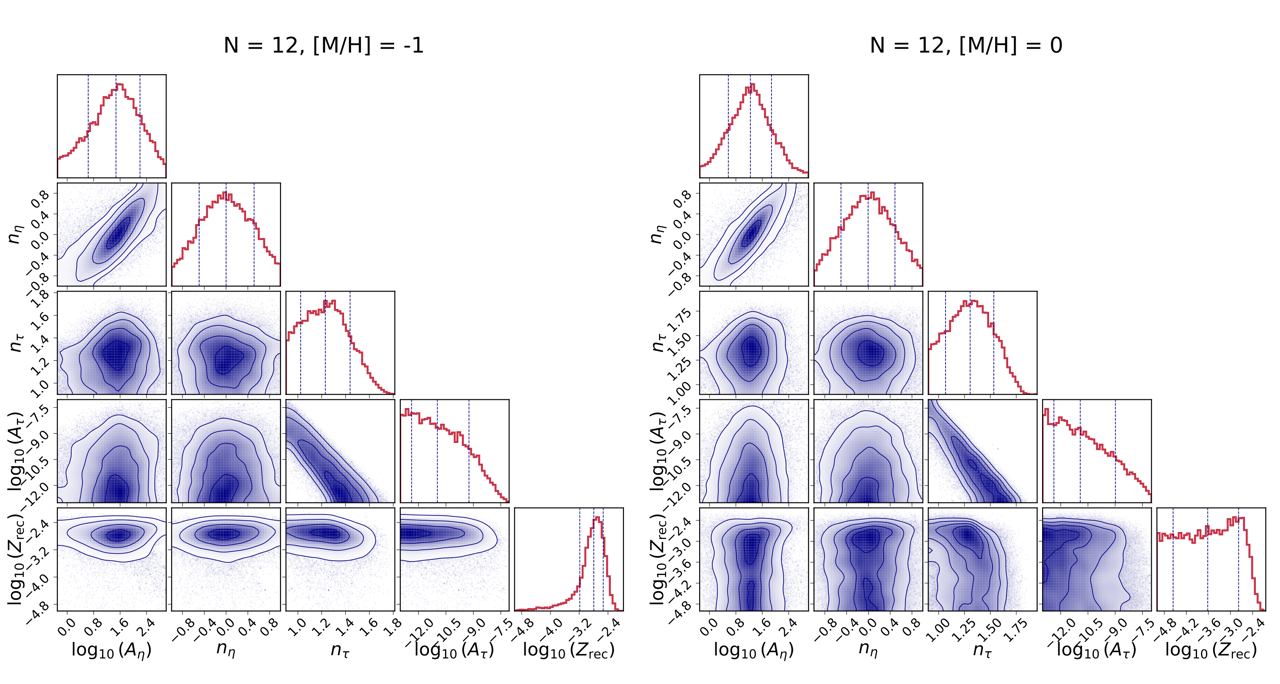}
\includegraphics[width=1.\textwidth]{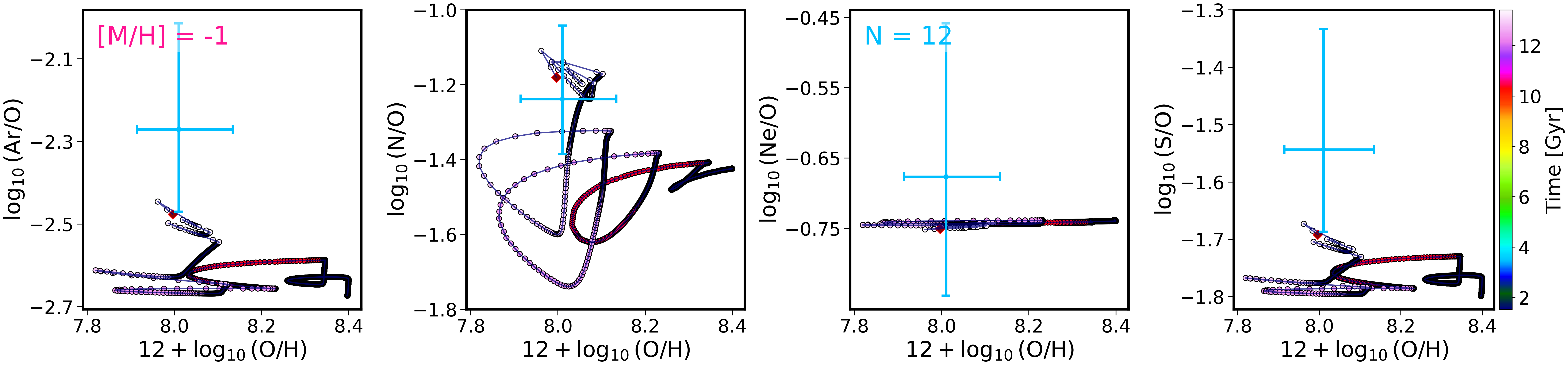}
\includegraphics[width=1.0\textwidth]{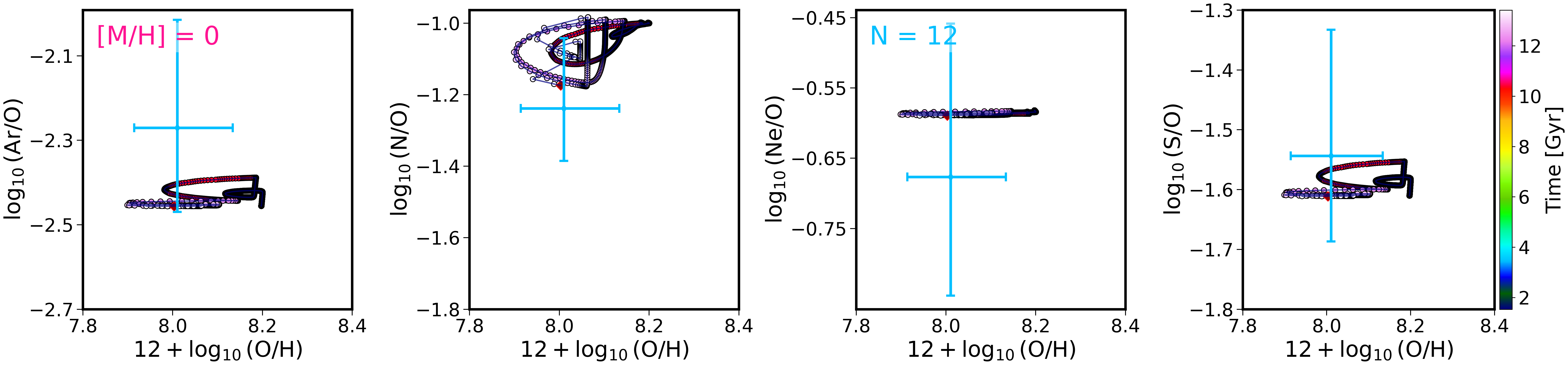}
\caption{Representative posterior constraints and best-fit enrichment trajectories for galaxy $N=12$. The top panel shows the posterior distribution of the five model parameters. The middle and bottom panels show the corresponding best-fit abundance-ratio tracks for the two yield-metallicity assumptions adopted in this work, $[\mathrm{M/H}] = -1$ and $0$, respectively. }  
\label{Results-Gal12}
\end{figure*}
%%%%%%%%%%%%%%%%%%%%%%%%%%%%%%%%%%%%%%%%%%%%%%%%%%%%

%%%%%%%%%%%%%%%%%%%%%%%%%%%%%%%%%%%%%%%%%%%%%%%%%%%%
\begin{figure*}[th!]
\center
\includegraphics[width=1.0\textwidth]{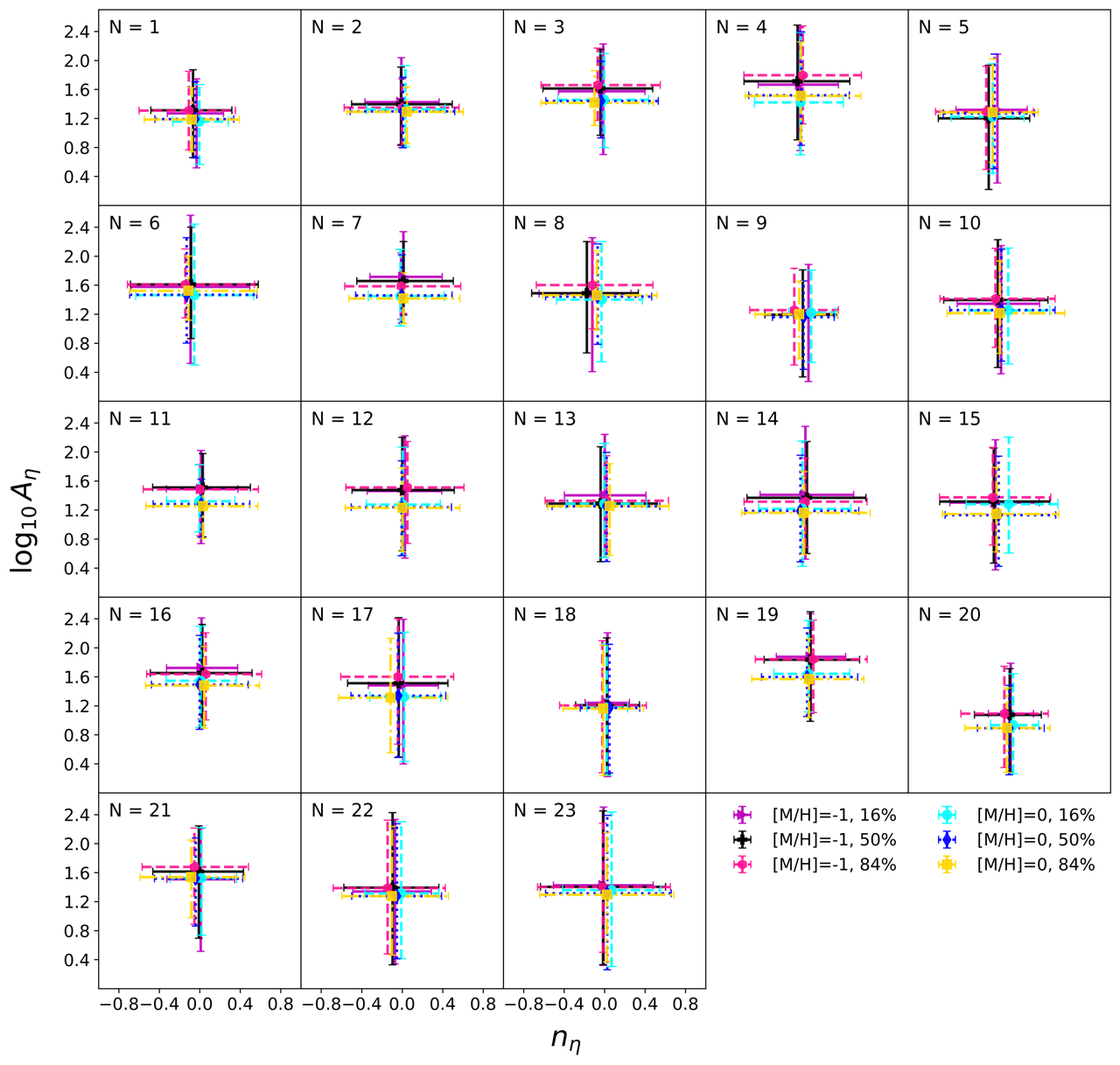}
\caption{Constraints on the outflow scaling parameters $n_{\eta}$ and $\log_{10}(A_{\eta})$ for the full sample of EELGs. Each panel corresponds to an individual galaxy, showing the median and 16/84 percentiles inferred from the MCMC analysis. Different markers denote results obtained using three SFH realizations from BAGPIPES (16\%, 50\%, 84\% percentiles), while colors distinguish between metallicity assumptions, $[\mathrm{M/H}] = -1$ and $0$. The inferred values show substantial overlap across SFH percentiles, indicating that the constraints are largely insensitive to the detailed bursty structure of the SFH. The dominant variation across galaxies is in the normalization $\log_{10}(A_{\eta})$, while $n_{\eta}$ remains confined to a relatively narrow range.}  
\label{neta-Aeta}
\end{figure*}
%%%%%%%%%%%%%%%%%%%%%%%%%%%%%%%%%%%%%%%%%%%%%%%%%%%%%

%%%%%%%%%%%%%%%%%%%%%%%%%%%%%%%%%%%%%%%%%%%%%%%%%%%%
\begin{figure*}[th!]
\center
\includegraphics[width=1.0\textwidth]{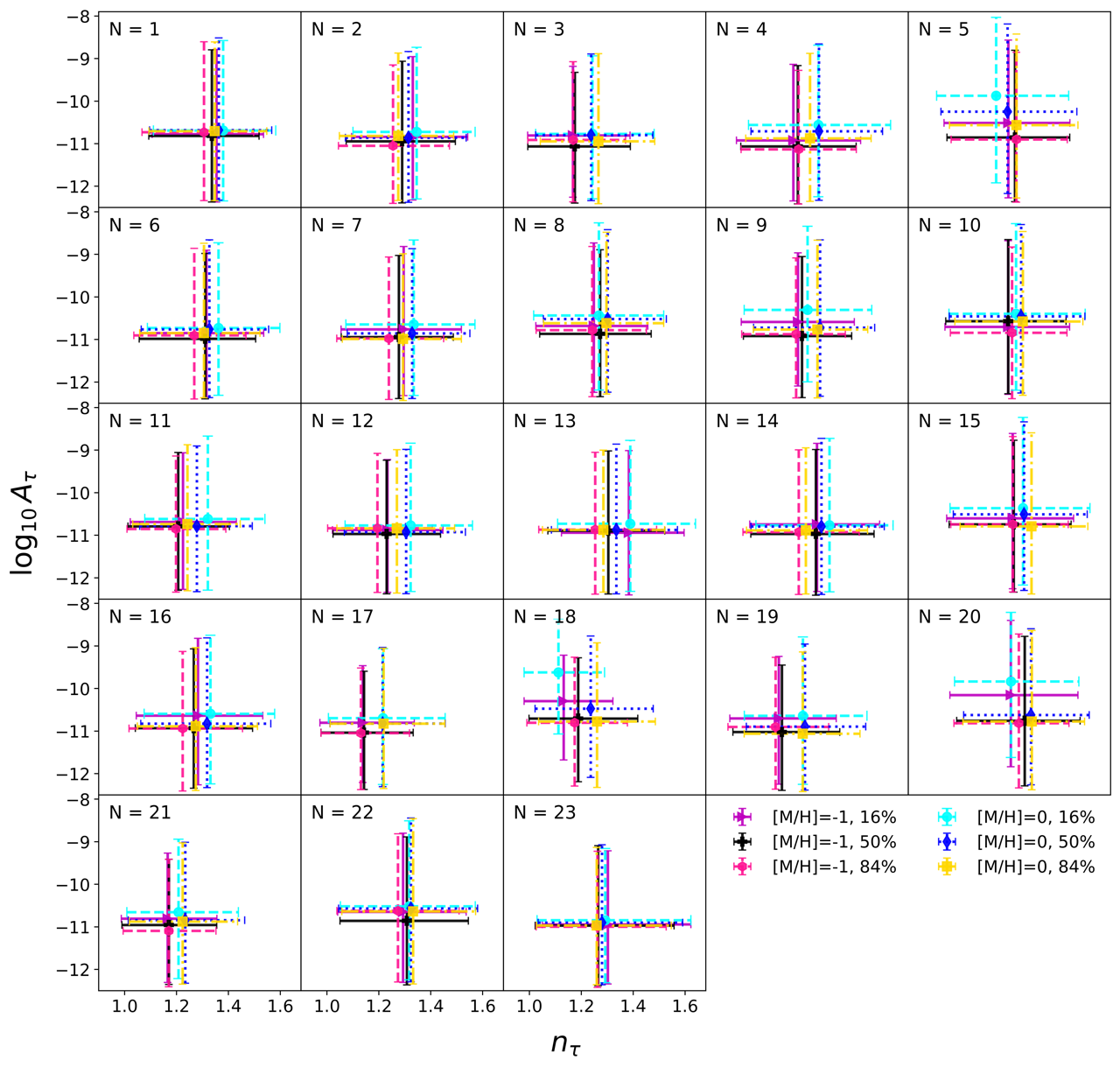}
\caption{Same as Figure \ref{neta-Aeta}, but for the star formation efficiency scaling parameters $n_{\tau}$ and $\log_{10}(A_{\tau})$. The inferred values of $n_{\tau}$ exhibit a high degree of uniformity across the sample, with strong overlap between different SFH realizations and metallicity assumptions. In contrast, the normalization $\log_{10}(A_{\tau})$ shows greater scatter and mild sensitivity to model assumptions. The clustering of $n_{\tau}$ highlights a common scaling behavior among these EELGs, despite their bursty star formation histories.}  
\label{ntau-Atau}
\end{figure*}
%%%%%%%%%%%%%%%%%%%%%%%%%%%%%%%%%%%%%%%%%%%%%%%%%%%%%

\subsection{Posterior constraints and trends}
\label{sec:mcmc-result}
Having described our MCMC Bayesian framework for parameter inference, we now present the key results obtained from this analysis and discuss the physical insights that emerge from the inferred parameter distributions.

Figure~\ref{Results-Gal12} presents a representative example of our MCMC results for Galaxy ID N=12. 
The top panels show the posterior parameter distributions from the corner plot, while the middle and bottom panels illustrate the resulting chemical enrichment trajectories obtained using two different stellar yield assumptions, corresponding to $[\mathrm{M/H}] = -1$ and $[\mathrm{M/H}] = 0$. 

The corner plots indicate that the inferred parameters governing the large-scale gas cycling of the galaxy, namely the mass-loading factor parameters and the depletion timescale, remain broadly consistent between the two yield sets. This behavior is expected, as these parameters primarily reflect global gas-regulation processes and should therefore depend only weakly on the adopted stellar yields. In contrast, the inferred metallicity of the recycled gas shifts noticeably between the two cases. This reflects a well-known degeneracy in chemical evolution models, in which the observed gas-phase metallicity depends on both the stellar metal yield and the metallicity of the inflowing or recycled gas. 
In the framework of gas-regulator models \citep[e.g.,][]{Lilly2013}, the equilibrium gas metallicity can be approximated as
\begin{equation}
Z_{\rm gas} \simeq \frac{y}{1+\eta} + Z_{\rm in},
\end{equation}
where $y$ is the stellar metal yield, $\eta$ is the mass-loading factor, and $Z_{\rm in}$ represents the metallicity of the inflowing or recycled gas. 
As a consequence, adopting more metal-poor stellar yields ($[\mathrm{M/H}] = -1$) requires comparatively less metal-poor recycled gas to reproduce the observed abundances, whereas more metal-rich stellar yields ($[\mathrm{M/H}] = 0$) require a correspondingly lower recycled-gas metallicity in order to match the same observational constraints.

This effect is clearly visible in the enrichment trajectories shown in the middle and bottom rows. 
Although the global gas-regulation parameters remain similar, the detailed paths followed in abundance space differ significantly depending on the adopted yield metallicity set. 
These results therefore highlight an important distinction: while the parameters governing the galaxy's gas cycling appear robustly constrained, the detailed chemical enrichment trajectory, and thus the interpretation of the metallicity evolution, remains sensitive to the underlying nucleosynthetic yield assumptions.

%The best-fit trajectories of the elemental abundance ratios for galaxies with two distinct metallicity yields ($[\mathrm{M/H}] = -1, 0$), along with the full results of the MCMC fits, are presented in Appendix \ref{sec:GCE} (Figures \ref{Time-evolutoin1}–\ref{Time-evolutoin23}) and Appendix \ref{sec:Corners} (Figures \ref{Corner1}–\ref{Corner23}), respectively. The trajectories reproduce the DESI data points within their associated uncertainties, while the corner plots demonstrate that the posterior distributions are well converged.

While the default Bayesian analysis assumes a fixed Type Ia supernova delay time of $\tau_{\mathrm{del}} = 2.0$ Gyr, we investigated the impact of changing delay time for Type Ia SNe and demonstrated that it does not have a significant contribution. Additionally, we adopt the AGB star yields exclusively from \cite{2011ApJS..197...17C} for this comparison.

Figure \ref{neta-Aeta} and Figure \ref{ntau-Atau} present the median and 16/84 percentiles of the posterior distributions in the $n_{\eta}$–$\log_{10}(A_{\eta})$ and $n_{\tau}$–$\log_{10}(A_{\tau})$ planes, respectively. In each panel, we show the MCMC results for three distinct realizations of the SFH, drawn from the BAGPIPES fits corresponding to the 16\%, 50\%, and 84\% percentiles, as well as for two different metallicity assumptions, $[\mathrm{M/H}] = -1$ and $0$. Overall, the inferred one-sigma ranges of these parameters are broadly consistent across both SFH realizations and metallicity choices, indicating that the chemical-evolution inference is relatively insensitive to these inputs.

For a given galaxy, the solutions obtained from the three SFH percentiles largely overlap, with shifts that are typically smaller than the associated uncertainties. This suggests that, despite the bursty and non-parametric nature of the BAGPIPES SFHs, the inferred chemical parameters are governed primarily by the integrated gas-processing history rather than by individual burst features. In the $n_{\eta}$–$\log_{10}(A_{\eta})$ plane, most of the variation across galaxies is driven by the normalization $\log_{10}(A_{\eta})$, while $n_{\eta}$ remains confined to a relatively narrow range around zero. A mild but recurring offset is observed between the two metallicity assumptions, with $[\mathrm{M/H}] = -1$ generally favoring slightly higher values of $\log_{10}(A_{\eta})$ compared to $[\mathrm{M/H}] = 0$, although this trend is not universal.

A stronger degree of uniformity is evident in the $n_{\tau}$–$\log_{10}(A_{\tau})$ plane, where the inferred values of $n_{\tau}$ cluster tightly across the full sample, largely independent of SFH percentile or metallicity choice. This indicates that the time dependence of the star formation efficiency timescale is remarkably consistent among these EELGs. As in the outflow case, the primary freedom lies in the normalization $\log_{10}(A_{\tau})$, which exhibits greater scatter and, in some galaxies, increased sensitivity to the adopted metallicity.

Taken together, these results indicate that variations in SFH percentile and metallicity primarily introduce moderate shifts in the normalization parameters, without significantly altering the preferred region of parameter space. Given the relatively uninformative priors adopted in the MCMC analysis, as presented in Equation \ref{priors}, the consistent clustering of the posteriors suggests that the data impose meaningful constraints on the chemical-evolution parameters. This points toward a scenario in which EELGs, despite their bursty star formation histories, occupy a broadly similar gas-regulation regime, with galaxy-to-galaxy differences driven mainly by normalization effects rather than by changes in the underlying functional dependencies. 

While our inferred values of $n_{\tau}$ are broadly consistent with the canonical Kennicutt–Schmidt scaling within uncertainties \citep{1959ApJ...129..243S,1963ApJ...137..758S,1998ApJ...498..541K}, we find that the majority of our constraints lie systematically toward the lower end of the reported range, $n \simeq 1.4 \pm 0.15$. In particular, the $1\sigma$ intervals of $n_{\tau}$ across our sample are typically offset toward smaller values, indicating a shallower effective scaling between gas content and star formation efficiency. This behavior is consistently observed across different SFH realizations and metallicity assumptions, suggesting that it is not driven by modeling choices but instead reflects an intrinsic property of the galaxies.
A natural interpretation is that the bursty star formation histories characteristic of these extreme emission line galaxies (EELGs) lead to departures from the quasi-equilibrium conditions underlying the classical Kennicutt–Schmidt relation. In such systems, rapid variability in gas inflow, feedback, and star formation introduces temporal offsets between the gas reservoir and the instantaneous star formation rate, breaking the tight, steady mapping assumed in equilibrium models. As a result, galaxies cycle through phases in which the gas content and star formation rate are out of sync—e.g., post-burst depletion or pre-burst gas buildup—so that the instantaneous relation is not single-valued. When averaged over these evolutionary cycles, this naturally leads to a shallower effective scaling, as reflected in the lower $n_{\tau}$ values we infer. This interpretation is further supported by the robustness of our results to SFH percentile choice, indicating that the effect is not tied to a particular realization of the bursty history but is instead a generic consequence of time-variable star formation. Our results therefore suggest that EELGs occupy a regime in which the Kennicutt–Schmidt relation remains broadly applicable, but with systematically lower effective slopes, highlighting the impact of bursty, non-equilibrium star formation on galactic scaling relations.

%%%%%%%%%%%%%%%%%%%%%%%%%%%%%%%%%%%%%%%%%%%%%%%%%%%%
\begin{figure*}[th!]
\center
\includegraphics[width=1.0\textwidth]{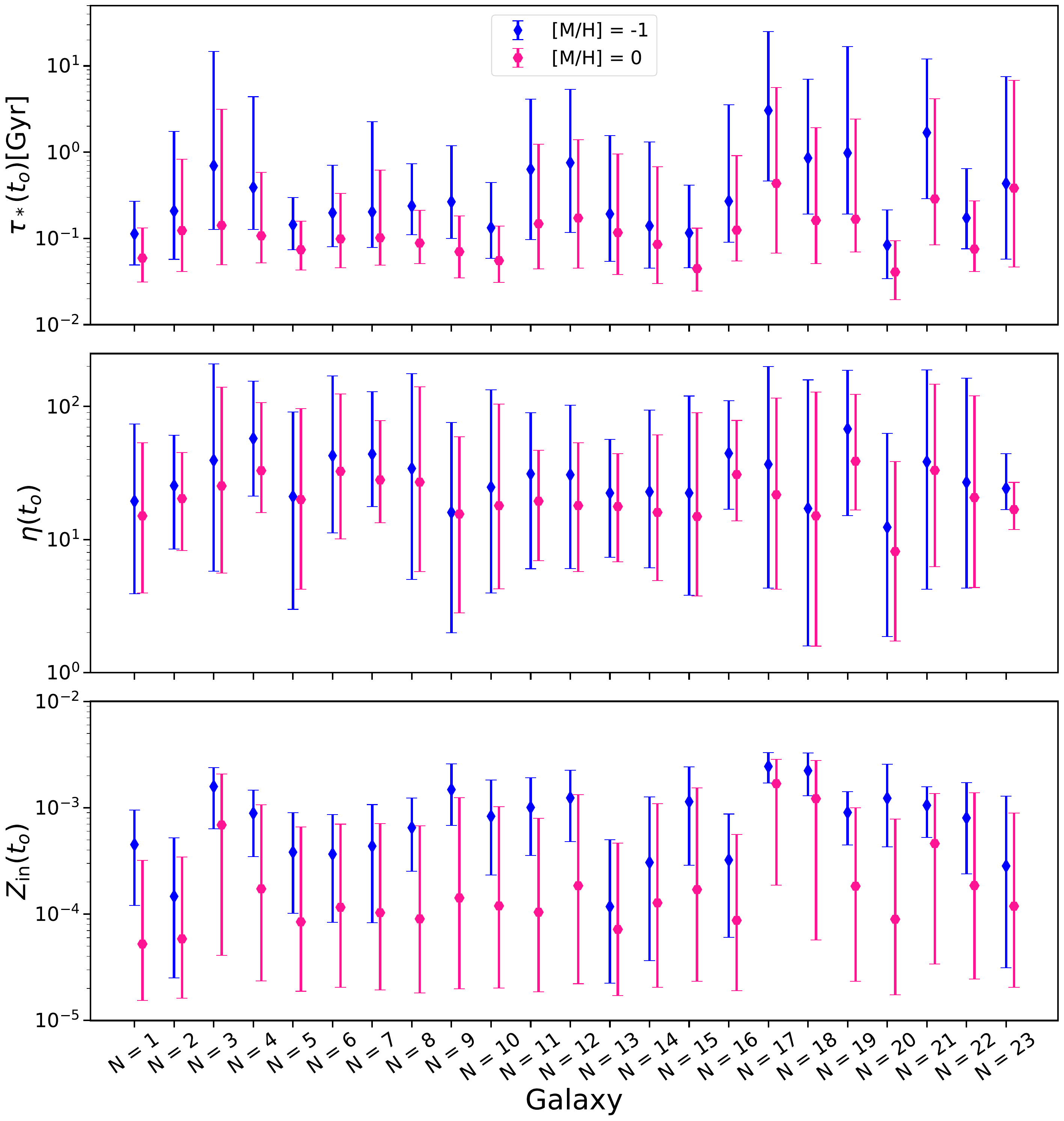}
\caption{Present-day ($t_{o}$) values of the star formation efficiency timescale $\tau_\star$, mass-loading factor $\eta$, and inflow metallicity $Z_{\rm in}$ for all galaxies in the EELG sample, derived from the pooled posterior distributions. Each panel shows the median and 16\%–84\% credible intervals obtained by combining the posterior samples from the three BAGPIPES SFH realizations (16\%, 50\%, 84\% percentiles). Results are shown for two metallicity assumptions, $[\mathrm{M/H}] = -1$ (blue) and $[\mathrm{M/H}] = 0$ (pink). All quantities are evaluated at the observed time and therefore reflect the most recent evolutionary phase of each galaxy, immediately prior to observation, given their bursty star formation histories (Figure~\ref{Fig:SFH-fit}). The inferred short depletion timescales and large mass-loading factors indicate highly efficient star formation and strong feedback-driven outflows, while the range of inflow metallicities, anchored by a primordial floor of $Z_{\rm prim} = 10^{-5}$, reflects a mixture of pristine accretion and recycled enrichment.}  
\label{Summary-tau-eta-Zin}
\end{figure*}
%%%%%%%%%%%%%%%%%%%%%%%%%%%%%%%%%%%%%%%%%%%%%%%%%%%%%
Finally, to derive physically interpretable quantities from the inferred parameter posteriors, we propagate the MCMC chains into time-dependent quantities for each galaxy from Equations \ref{Tau-form}-\ref{z-infall} . For a given object, we load the posterior samples obtained from the three BAGPIPES-based SFH realizations corresponding to the 16\%, 50\%, and 84\% percentiles. To ensure equal weighting across these SFH realizations, we resample each posterior chain to the same number of samples and construct an equal-weight mixture. For each SFH realization, we reconstruct the star formation history as a continuous function of cosmic time by applying the same preprocessing used in our fitting pipeline: truncation at $t \leq t_{o} - 0.001,\mathrm{Gyr}$, smoothing via a Gaussian kernel, and interpolation onto a common time grid. This guarantees consistency between the inferred parameters and the time-dependent quantities derived from them.

Using these posterior samples and reconstructed SFHs, we compute the time evolution of the star formation efficiency timescale $\tau_\star(t)$ and the mass-loading factor $\eta(t)$ in a fully vectorized manner. At each time step, these quantities are evaluated directly from the model parameterization for every posterior draw, yielding distributions $\tau_\star(t)$ and $\eta(t)$ that incorporate the full parameter uncertainties. In addition, we compute the inflow metallicity $Z_{\rm in}(t)$ following the prescription introduced in Equations~\ref{z-infall} and \ref{S-function}, in which the transition from primordial to recycled enrichment is governed by the SFH itself. The timing of this transition is determined from the cumulative SFH, while the normalization is set by the posterior distribution of the recycled metallicity. Elemental inflow abundances are then obtained by scaling the total inflow metallicity according to solar abundance ratios.

At the end, we combine the posterior samples from the three SFH realizations by stacking them to construct a pooled distribution for each galaxy. From these pooled samples, we compute the median and 16\%–84\% credible intervals as a function of time for all derived quantities. The present-day values, evaluated at $t_{o}$, are extracted from the final time step and used for comparison across galaxies and between different metallicity assumptions, $[\mathrm{M/H}] = -1$ and $0$. This procedure allows us to robustly propagate both parameter uncertainties and SFH-driven systematics into physically meaningful constraints on the current gas depletion timescale, mass-loading factor, and inflow metallicity of each system.

Figure \ref{Summary-tau-eta-Zin} shows the inferred present-day values of the star formation efficiency timescale $\tau_\star(t_{o})$, the mass-loading factor $\eta(t_{o})$, and the inflow metallicity $Z_{\rm in}(t_{o})$ for all galaxies in our EELG sample, derived from the pooled posterior distributions. As seen in the top panel, the depletion timescales are systematically short, typically $\tau_\star \sim 0.05$–$10,\mathrm{Gyr}$, with the majority of systems clustered below $\sim 0.5,\mathrm{Gyr}$. These values correspond to the conditions at the observed time and, given the bursty star formation histories shown in Figure~\ref{Fig:SFH-fit}, primarily reflect the most recent phase of star formation immediately preceding the observation. The comparison between the two metallicity assumptions shows a mild but systematic shift, with $[\mathrm{M/H}] = -1$ generally yielding slightly larger $\tau_\star$ values than $[\mathrm{M/H}] = 0$, although the two cases remain broadly consistent within uncertainties. The short depletion times inferred here are significantly lower than those typically observed in more massive, main-sequence star-forming galaxies, reinforcing the interpretation that EELGs are undergoing intense, transient star-forming episodes.

The depiction of the mass-loading factor $\eta(t_{o})$ in the middle panel reveals that these systems are also characterized by elevated mass-loading factors, typically $\eta \sim 10$–$100$, and in some cases extending to even higher values. Such large $\eta$ values imply that outflows are dynamically important and likely play a dominant role in regulating the gas content of these galaxies. As with $\tau_\star$, these estimates correspond to the instantaneous conditions at $t_{o}$ and therefore trace the current feedback state of the system during its most recent burst. The combination of short depletion times and high mass-loading factors places EELGs in a regime of rapid gas cycling, where inflows, star formation, and feedback operate on comparably short timescales. This behavior is consistent with expectations for low-mass, bursty systems, in which stellar feedback can efficiently expel gas and drive strong fluctuations in the star formation rate.

Finally, the bottom panel shows that the inferred inflow metallicities span a wide range, $Z_{\rm in}(t_{o}) \sim 10^{-5}$ – $5 \times 10^{-3}$, with a clear offset between the two metallicity assumptions. In our model, the primordial inflow metallicity is fixed at $Z_{\rm prim} = 10^{-5}$, and the inferred values therefore reflect a combination of pristine inflow and enriched recycled gas. The $[\mathrm{M/H}] = -1$ models consistently favor higher inflow metallicities compared to the $[\mathrm{M/H}] = 0$ case, reflecting the interplay between the assumed enrichment baseline and the recycled component of the inflowing gas. The broad range of $Z_{\rm in}$ indicates that these galaxies are not accreting purely pristine material, but instead experience significant mixing between low-metallicity inflows and previously enriched gas. Taken together, these results support a picture in which EELGs occupy a highly non-equilibrium regime characterized by short gas depletion times, strong outflows, and time-variable enrichment, driven by their bursty star formation histories.
%%%%%%%%%%%%%%%%%%%%%%%%%%%%%%%%%%%%%%%%%%%%%%%%%%%%
\begin{table} [!htbp] 
\centering
\begin{tabular}{|l|c|c|c|c|c|r|} 
		\hline 
		parameters &  $n_{\tau}$  &
        $\log_{10}{(A_{\tau})}$ & 
         $n_{\eta}$  &
         $\log_{10}{(A_{\eta})}$  & 
        $\log_{10}{(Z_{\rm rec})}$
        \\
        \hline
		default & 1.2 & -11.0 & 0.0 & 1.5 & -2.8
        \\    
        \hline
       	\end{tabular}
	\caption{Default values chosen for our desired galactic parameters, ($n_{\tau}$,
        $\log_{10}{(A_{\tau})}$, 
         $n_{\eta}$,
         $\log_{10}{(A_{\eta})}$, 
        $\log_{10}{(Z_{\rm rec})}$), used in making Figure \ref{Abundance-behavior}. }
\label{Default_Values1}
\end{table}
%%%%%%%%%%%%%%%%%%%%%%%%%%%%%%%%%%%%%%%%%%%%%%%%%%%%
\subsection{A schematic picture of the role of varying galactic parameters}
\label{sec:sxhematic-View}
\begin{figure*}[th!]
\center
\includegraphics[width=1.0\textwidth]{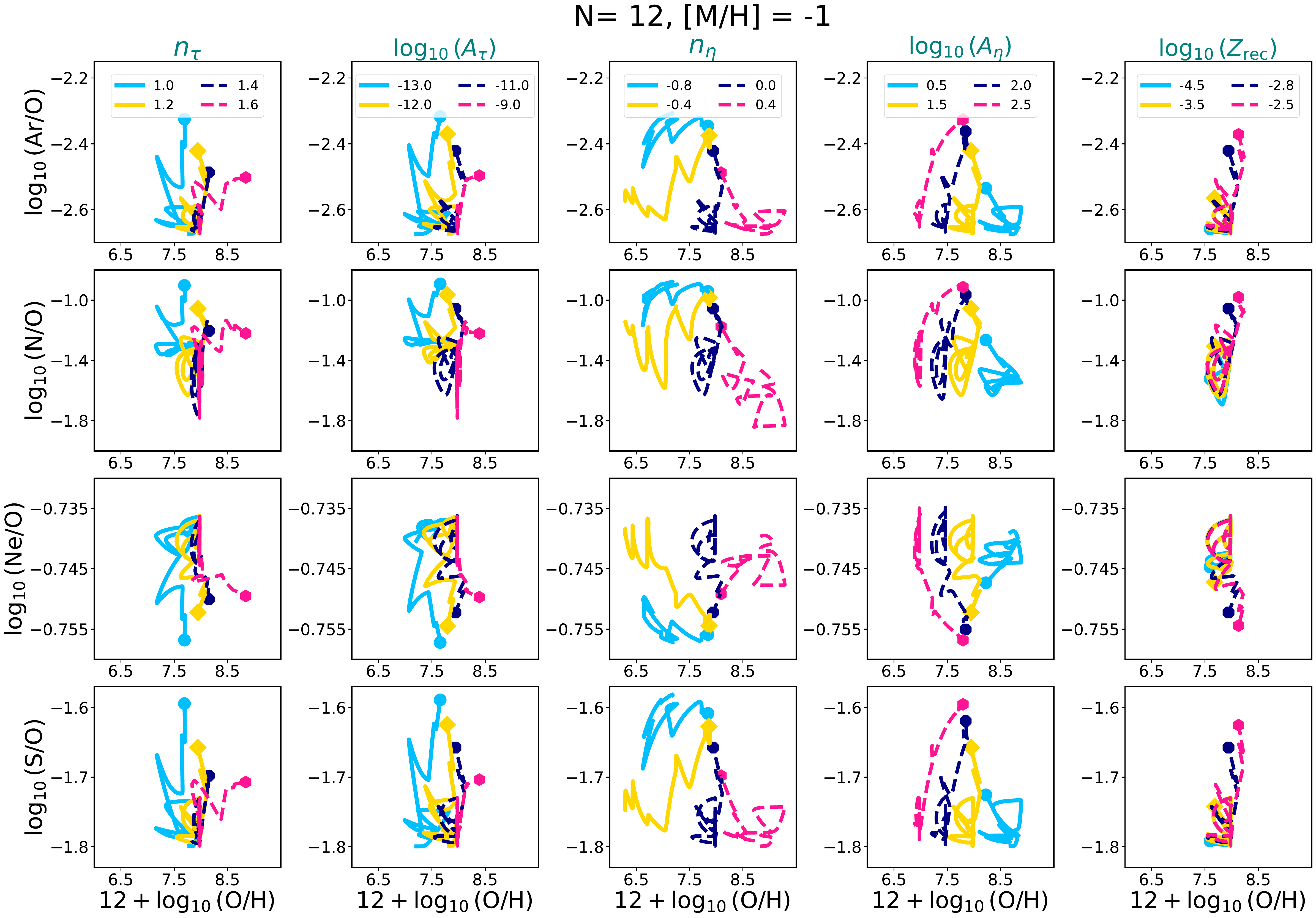}
\caption{Sensitivity of chemical abundance tracks to individual model parameters for a representative galaxy ($N=12$). Each panel shows $\log_{10}(\mathrm{X/O})$ versus $12+\log_{10}(\mathrm{O/H})$, varying one parameter at a time—$n_{\tau}$, $\log_{10}{(A_{\tau})}$, $n_{\eta}$, $\log_{10}{(A_{\eta})}$, and $\log_{10}{(Z_{\rm rec})}$—while fixing the others at their fiducial values (Table~\ref{Default_Values1}). Variations in $n_{\tau}$ and $\log_{10}{(A_{\tau})}$ primarily drive diagonal shifts associated with enrichment timescales, whereas $\log_{10}{(A_{\eta})}$ and $n_{\eta}$ introduce stronger vertical and diagonal displacements due to metal loss. The parameter $\log_{10}{(Z_{\rm rec})}$ produces an approximately vertical shift, reflecting its role in setting the inflow metallicity. Differences among elements are evident, with $\mathrm{N/O}$ showing the strongest response and $\alpha$-element ratios exhibiting comparatively weaker variations.}  
\label{Abundance-behavior}
\end{figure*}

To build physical intuition for how individual model parameters shape the observed abundance patterns, we perform a controlled set of one-dimensional variations around a fiducial model for a representative system (here, galaxy $N=12$). In this experiment, each parameter—$n_{\tau}$, $\log_{10}{(A_{\tau})}$, $n_{\eta}$, $\log_{10}{(A_{\eta})}$, and $\log_{10}{(Z_{\rm rec})}$—is varied independently over a predefined grid while all other parameters are fixed at their default values (Table~\ref{Default_Values1}). The resulting evolutionary tracks in $\log_{10}(\mathrm{X/O})$ ($X \in \left[ \mathrm{Ar}, \mathrm{N}, \mathrm{Ne}, \mathrm{S} \right] $) versus $12+\log_{10}(\mathrm{O/H})$ space are shown in Figure~\ref{Abundance-behavior}. 

From these tracks, we identify several robust trends. Variations in the star formation efficiency parameters ($n_{\tau}$ and $\log_{10}{(A_{\tau})}$) primarily alter the evolutionary progression along the tracks, producing diagonal shifts in abundance space that reflect changes in the rate and timing of chemical enrichment. Increasing these parameters generally drives the system toward higher $12+\log_{10}(\mathrm{O/H})$, accompanied by modest changes in $\log_{10}(\mathrm{X/O})$ that depend on the element. In contrast, the outflow parameters ($n_{\eta}$ and $\log_{10}{(A_{\eta})}$) regulate the efficiency of metal removal and therefore introduce stronger vertical and diagonal shifts. Increasing $\log_{10}{(A_{\eta})}$ shifts the tracks toward lower metallicities and typically higher $\log_{10}(\mathrm{X/O})$, while variations in $n_{\eta}$ modify both the slope and curvature of the evolutionary tracks. 

The recycled inflow metallicity, $\log_{10}{(Z_{\rm rec})}$, produces a displacement across all abundance ratios of X/O, acting as an enrichment baseline with similar impact on the horizontal evolution in $12+\log_{10}(\mathrm{O/H})$.

The response of abundance ratios also depends on nucleosynthetic origin. Ratios involving elements with delayed or secondary enrichment channels, such as $\mathrm{N/O}$, exhibit the rather strong sensitivity to parameter variations. In contrast, ratios among $\alpha$-elements—such as $\mathrm{Ne/O}$, $\mathrm{Ar/O}$, and $\mathrm{S/O}$—are generally more stable, as these elements are produced in similar sites and track each other more closely. However, $\mathrm{Ne/O}$ shows the weakest variation overall, while $\mathrm{Ar/O}$ and $\mathrm{S/O}$ display moderate sensitivity, particularly to outflows and enrichment history, indicating that even among $\alpha$-elements, the response is not strictly identical.

\begin{figure*}[th!]
\center
\includegraphics[width=0.8\textwidth]{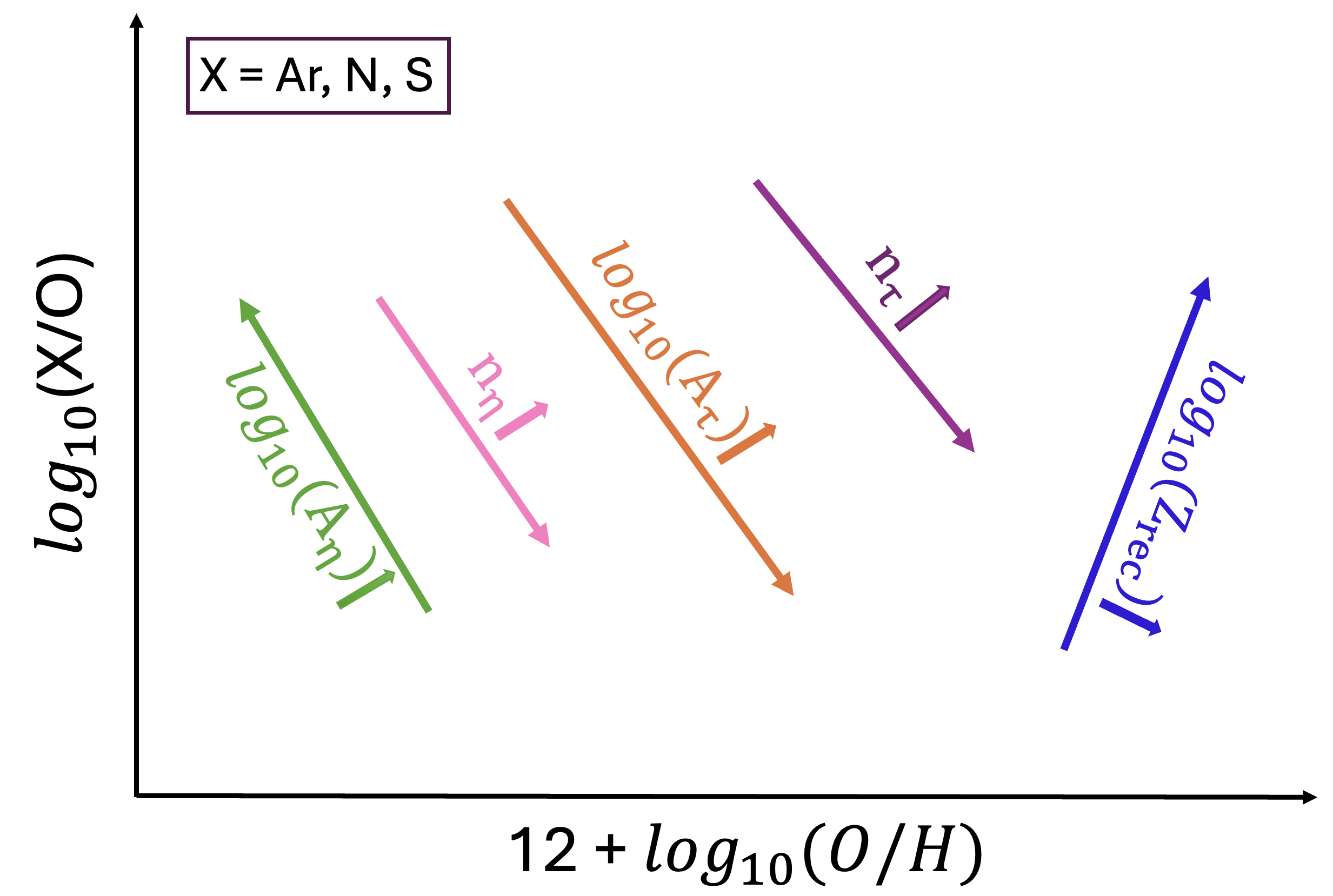}
\includegraphics[width=0.8\textwidth]{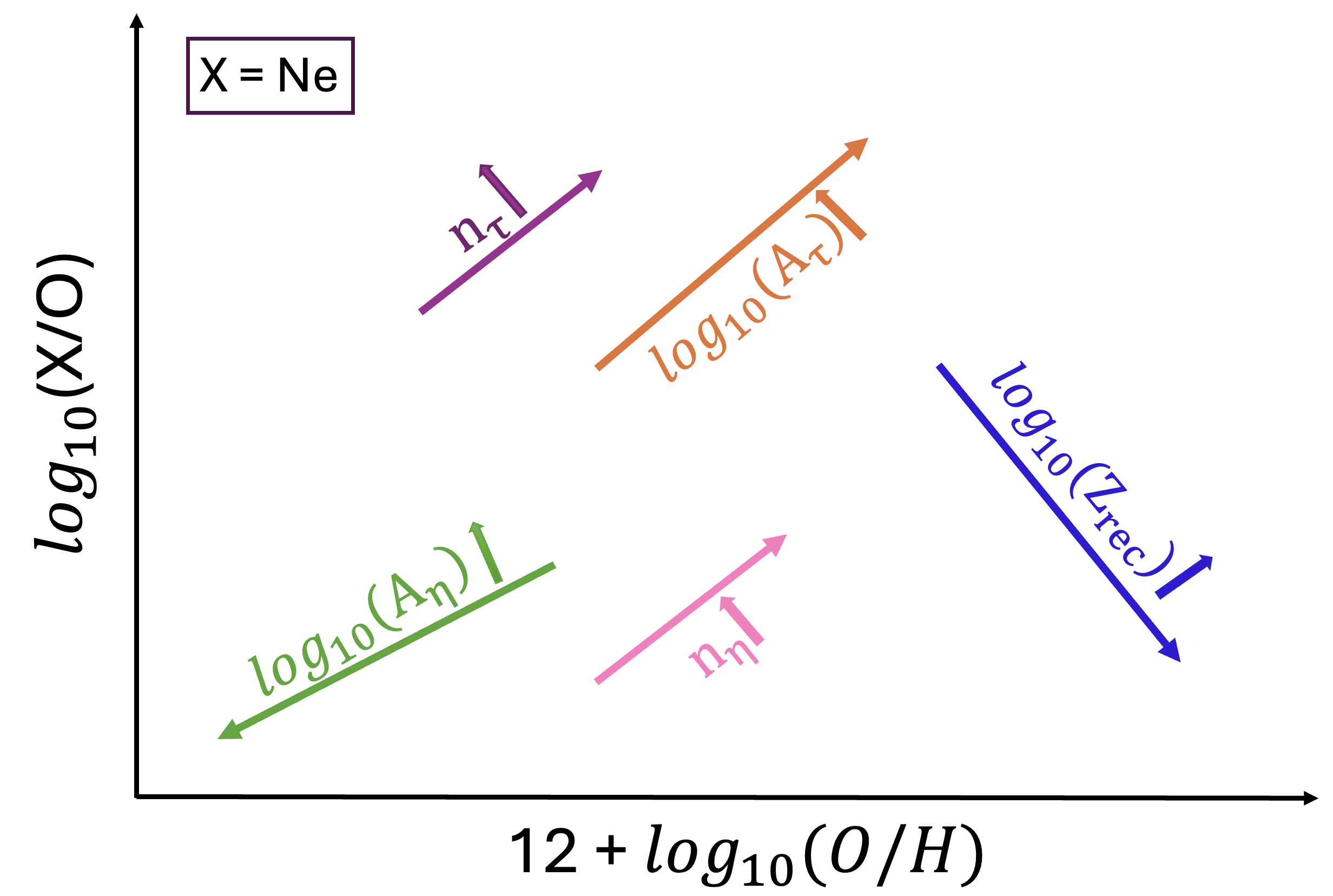}
\caption{Schematic summary of the parameter dependencies shown in Figure~X. Arrows indicate the direction of change in $\log_{10}(\mathrm{X/O})$ versus $12+\log_{10}(\mathrm{O/H})$ as each parameter increases. Variations in $n_{\tau}$ and $\log_{10}{(A_{\tau})}$ produce diagonal shifts linked to enrichment timescales, while $\log_{10}{(A_{\eta})}$ and $n_{\eta}$ introduce vertical and diagonal displacements associated with outflows. The parameter $\log_{10}{(Z_{\rm rec})}$ produces an approximately vertical shift. The bottom panel highlights the weaker response of $\mathrm{Ne/O}$ compared to other abundance ratios.} 
\label{Carton-behavior}
\end{figure*}

To summarize these behaviors in a more compact and intuitive form, Figure~\ref{Carton-behavior} presents a schematic representation of the dominant trends identified in the grid exploration. This cartoon is not an independent model, but rather a qualitative condensation of the results shown in Figure~\ref{Abundance-behavior}, where the arrows indicate the direction of change in $12+\log_{10}(\mathrm{O/H})$ and $\log_{10}(\mathrm{X/O})$ as each parameter increases. In this framework, variations in $n_{\tau}$ and $\log_{10}{(A_{\tau})}$ primarily drive diagonal evolution associated with enrichment timescales, while $\log_{10}{(A_{\eta})}$ and $n_{\eta}$ introduce vertical and diagonal shifts linked to metal loss. The parameter $\log_{10}{(Z_{\rm rec})}$ produces more  vertical displacement compared to other parameters, reflecting its role in setting the baseline metallicity of inflowing gas.

Taken together, these results provide a unified physical picture of how chemical abundance patterns arise in EELGs. Different parameters leave distinct directional signatures in abundance space, while degeneracies emerge when multiple parameters produce similar net displacements. This framework not only explains the structure of the posterior distributions obtained from our MCMC analysis, but also reinforces the interpretation of EELGs as systems governed by bursty star formation, strong outflows, and time-variable enrichment operating in a non-equilibrium regime.

%%%%%%%%%%%%%%%%%%%%%%%%%%%%%%%%%%%%%%%%%%%%%%%%%%%%
%%%%%%%%%%%%%%%%%%%%%%%%%%%%%%%%%%%%%%%%%%%%%%%%%%%%

%%%%%%%%%%%%%%%%%%%%%%%%%%%%%%%%%%%%%%%%%%%%%%%%%%%%
%%%%%%%%%%%%%%%%%%%%%%%%%%%%%%%%%%%%%%%%%%%%%%%%%%%%

\subsection{Comparison to literature}
\label{comparison-literature}
Recent studies of low-mass star-forming galaxies have emphasized that these systems are gas-rich, metal-poor, and strongly affected by stellar feedback, leading to highly time-variable, non-equilibrium evolution \citep{2019ApJ...874...93B}. A similar picture has emerged for galaxies at higher redshifts, where rapid gas accretion, bursty star formation, and strong outflows are thought to drive chemical evolution in a non-steady manner \citep{2025A&A...697A..89C,2025MNRAS.541L..71I,2025A&A...696A..87C}. In this context, our EELG sample is fully consistent with this broader framework: the short present-day depletion timescales, large mass-loading factors, and wide range of inflow metallicities inferred in our analysis all point toward rapid gas cycling rather than quasi-equilibrium growth. Given the bursty star formation histories shown in Figure~\ref{Fig:SFH-fit}, our results suggest that the observed abundance patterns primarily reflect the most recent phase of star formation and feedback, reinforcing the interpretation that EELGs evolve through episodic enrichment events rather than smooth chemical evolution.

This behavior should not be interpreted as a strict physical degeneracy. In particular, the recycled inflow metallicity, $\log_{10}{(Z_{\rm rec})}$, acts as a baseline enrichment term, shifting abundance ratios primarily in the vertical direction, whereas the outflow normalization, $\log_{10}{(A_{\eta})}$, regulates the retention and removal of metals, introducing both vertical and diagonal changes in the abundance tracks. Although these parameters can lead to partially overlapping signatures in $\log_{10}(\mathrm{X/O})$ versus $12+\log_{10}(\mathrm{O/H})$ space, their physical roles remain distinct. This distinction is important, as it connects to the well-known degeneracy between stellar yields and inflow metallicity, where variations in metal production efficiency can be compensated by changes in the composition of accreted gas \citep{2017ApJ...835..128C}. Our results therefore provide a concrete, parameter-level illustration of how such degeneracies manifest in observable abundance space, and demonstrate the importance of multi-element analyses in disentangling them.

Another important outcome of our analysis is the differential sensitivity of abundance ratios to model parameters. We find that $\mathrm{N/O}$ exhibits the strongest variation, reflecting its dependence on delayed enrichment channels, while $\alpha$-element ratios such as $\mathrm{Ne/O}$, $\mathrm{Ar/O}$, and $\mathrm{S/O}$ are comparatively more stable. In particular, $\mathrm{Ne/O}$ shows the weakest response across all parameter variations, remaining nearly constant compared to other ratios. This behavior is consistent with observational studies that identify $\mathrm{Ne/O}$ as a relatively stable reference abundance ratio \citep{2024ApJ...968...98A}. In contrast, $\mathrm{Ar/O}$ and $\mathrm{S/O}$ display moderate sensitivity to both outflows and enrichment history, indicating that even among $\alpha$-elements, the response to galactic processes is not identical.

Taken together, these results highlight three key physical insights: (i) EELGs operate in a burst-driven, non-equilibrium regime characterized by rapid gas cycling, (ii) chemical abundance patterns encode distinct signatures of star formation efficiency, outflows, and inflow metallicity, and (iii) multi-element abundance analyses are essential for disentangling these effects. These conclusions place our work in direct continuity with previous studies of dwarf galaxies and galactic chemical evolution, while extending them by providing a unified, parameter-level interpretation of abundance trends in systems with bursty star formation histories.

\section{Conclusion}
\label{sec:Conclusion}

In this work, we use DESI DR1 observations to unravel the physical processes governing chemical enrichment in extreme emission-line galaxies, revealing a population shaped by burst-driven, non-equilibrium evolution. We construct a sample of galaxies with high-quality spectroscopic measurements, requiring the simultaneous detection of 19 ionic species with signal-to-noise ratios $\mathrm{S/N} \geq 4$, stellar masses above $M_* \geq10^7~M_\odot$, and large equivalent widths in H$\alpha$ and [O III] $\lambda5007$ ($\mathrm{EW} \geq 500\text{\AA}$). This selection yields 24 galaxies, for which we perform spectral energy distribution fitting using BAGPIPES. We find that 23 of these systems are well described by non-parametric star formation histories, all exhibiting clear signatures of recent, intense star formation episodes. The use of flexible, non-parametric SFHs minimizes biases associated with assumed functional forms and allows us to robustly capture the bursty nature of these systems.

Focusing on the 23 well-fit galaxies, we develop a Bayesian framework to model their chemical evolution. Using a single-zone formalism, we evolve each system forward in time to the observed epoch and compute abundance ratios for O, N, Ne, S, and Ar, which probe a range of nucleosynthetic channels and enrichment timescales. We parameterize the time dependence of the star formation efficiency and outflow mass-loading factor through flexible functional forms, and include a physically motivated prescription for the metallicity of inflowing gas. In this model, the inflow metallicity evolves from a primordial baseline ($Z_{\rm prim} = 10^{-5}$) to an enriched, recycled component, with the transition governed by the cumulative star formation history of the system. This time-dependent formulation allows us to consistently connect early, low-metallicity accretion with later recycling of enriched gas, naturally incorporating the extended star formation histories inferred from the data. By forward-modeling the chemical evolution and comparing directly to the observed abundance ratios, we constrain the posterior distributions of these parameters using MCMC sampling.

Our results reveal a consistent physical picture in which EELGs evolve in a burst-driven, non-equilibrium regime. The inferred depletion timescales are short and the mass-loading factors are large, indicating rapid gas consumption and efficient feedback-driven outflows. The parameters governing the time dependence of the depletion timescale and mass-loading factor show significant consistency across the sample, and remain robust to variations in the adopted SFH (16th, 50th, and 84th percentiles) and stellar metallicity ($[\mathrm{M/H}] = -1, 0$). In contrast, the recycled inflow metallicity exhibits greater sensitivity to these assumptions, reflecting its role as a baseline enrichment term and its connection to the degeneracy between metal production and inflow composition. We further find that the inferred scaling of the depletion timescale is broadly consistent with the Kennicutt–Schmidt relation, but tends toward slightly lower effective slopes, suggesting that bursty star formation leads to departures from the equilibrium scaling observed in more massive systems.

A central outcome of this work is the identification of how individual physical parameters shape abundance patterns. Through controlled parameter variations and schematic representations, we show that the star formation efficiency parameters primarily regulate the evolutionary progression in abundance space, while the outflow parameters govern the efficiency with which enriched gas is expelled from the system, thereby modulating the normalization of abundance ratios. The inflow metallicity acts as a baseline enrichment term, producing approximately vertical shifts in abundance space without strongly altering the evolutionary trajectory. Importantly, the sensitivity of abundance ratios depends on their nucleosynthetic origin. We find that $\mathrm{N/O}$ exhibits the strongest variation across the parameter space, reflecting its dependence on delayed enrichment channels and its sensitivity to both star formation history and gas flows. In contrast, $\mathrm{Ne/O}$ remains remarkably stable under all parameter variations, consistent with its origin in $\alpha$-element nucleosynthesis and its close tracking of oxygen. Intermediate behavior is observed for $\mathrm{Ar/O}$ and $\mathrm{S/O}$, which, while also $\alpha$-elements, show moderate sensitivity to outflows and enrichment history. These trends demonstrate that different abundance ratios carry complementary information, with stable ratios such as $\mathrm{Ne/O}$ serving as reference baselines and more sensitive ratios such as $\mathrm{N/O}$ providing strong diagnostic power. These effects leave distinct directional signatures, while similarities in their impact can arise when different parameters produce comparable net displacements, particularly in the vertical normalization of abundance ratios. This behavior highlights the importance of multi-element abundance measurements in disentangling the contributions of star formation, outflows, and inflow enrichment.

Finally, we emphasize both the strengths and limitations of our approach. While we are able to place meaningful constraints on the time-dependent star formation efficiency and outflow mass-loading factor, the detailed trajectories of chemical enrichment remain sensitive to assumptions about stellar yields and inflow metallicity. Nevertheless, the robustness of our key results across different SFH realizations and metallicity assumptions demonstrates that the primary physical conclusions—namely, burst-driven evolution, rapid gas cycling, and strong feedback—are not driven by modeling choices.

More broadly, our results have direct implications for models of galaxy formation in a cosmological context, particularly for low-mass, extreme emission-line systems. The prevalence of short depletion timescales, large mass-loading factors, and time-dependent inflow metallicities in our sample supports a picture in which EELGs evolve through highly time-variable, non-equilibrium baryon cycling, rather than steady-state growth. This behavior is consistent with expectations from modern cosmological simulations of low-mass galaxies, in which feedback-driven gas flows regulate star formation and chemical enrichment. At the same time, we emphasize that our conclusions are based on a sample of extreme, burst-dominated systems, and may not directly generalize to more typical star-forming galaxies. Extending this framework to broader galaxy populations, including systems with less extreme star formation activity, will be essential for establishing how widely these non-equilibrium processes govern galaxy evolution.

Our analysis thus establishes a coherent framework for interpreting chemical abundance patterns in EELGs, and demonstrates the power of combining large spectroscopic datasets with flexible chemical evolution modeling to probe the physical processes that shape galaxy evolution in chemically young, dynamically evolving systems.

\section*{Acknowledgments}

We are deeply grateful to Daniel Eisenstein for extensive guidance and substantial contributions to this work. We also thank Sophia Kempe and James Johnson for valuable discussions. The Legacy Surveys consist of three individual and complementary projects: the Dark Energy Camera Legacy Survey (DECaLS; Proposal ID \#2014B-0404; PIs: David Schlegel and Arjun Dey), the Beijing-Arizona Sky Survey (BASS; NOAO Prop. ID \#2015A-0801; PIs: Zhou Xu and Xiaohui Fan), and the Mayall z-band Legacy Survey (MzLS; Prop. ID \#2016A-0453; PI: Arjun Dey). DECaLS, BASS and MzLS together include data obtained, respectively, at the Blanco telescope, Cerro Tololo Inter-American Observatory, NSF’s NOIRLab; the Bok telescope, Steward Observatory, University of Arizona; and the Mayall telescope, Kitt Peak National Observatory, NOIRLab. Pipeline processing and analyses of the data were supported by NOIRLab and the Lawrence Berkeley National Laboratory (LBNL). The Legacy Surveys project is honored to be permitted to conduct astronomical research on Iolkam Du’ag (Kitt Peak), a mountain with particular significance to the Tohono O’odham Nation.

NOIRLab is operated by the Association of Universities for Research in Astronomy (AURA) under a cooperative agreement with the National Science Foundation. LBNL is managed by the Regents of the University of California under contract to the U.S. Department of Energy.

K.B. is supported by an NSF Astronomy and Astrophysics Postdoctoral Fellowship under award AST-2303858. JAAT acknowledges support from the Simons Foundation and \emph{JWST} program 3215. Support for program 3215 was provided by NASA through a grant from the Space Telescope Science Institute, which is operated by the Association of Universities for Research in Astronomy, Inc., under NASA contract NAS 5-03127.

\appendix

\section{Optical Image Gallery}
\label{sec:appendix-galleries}

\begin{figure*}[th!]
\center
\includegraphics[width=0.9\textwidth]{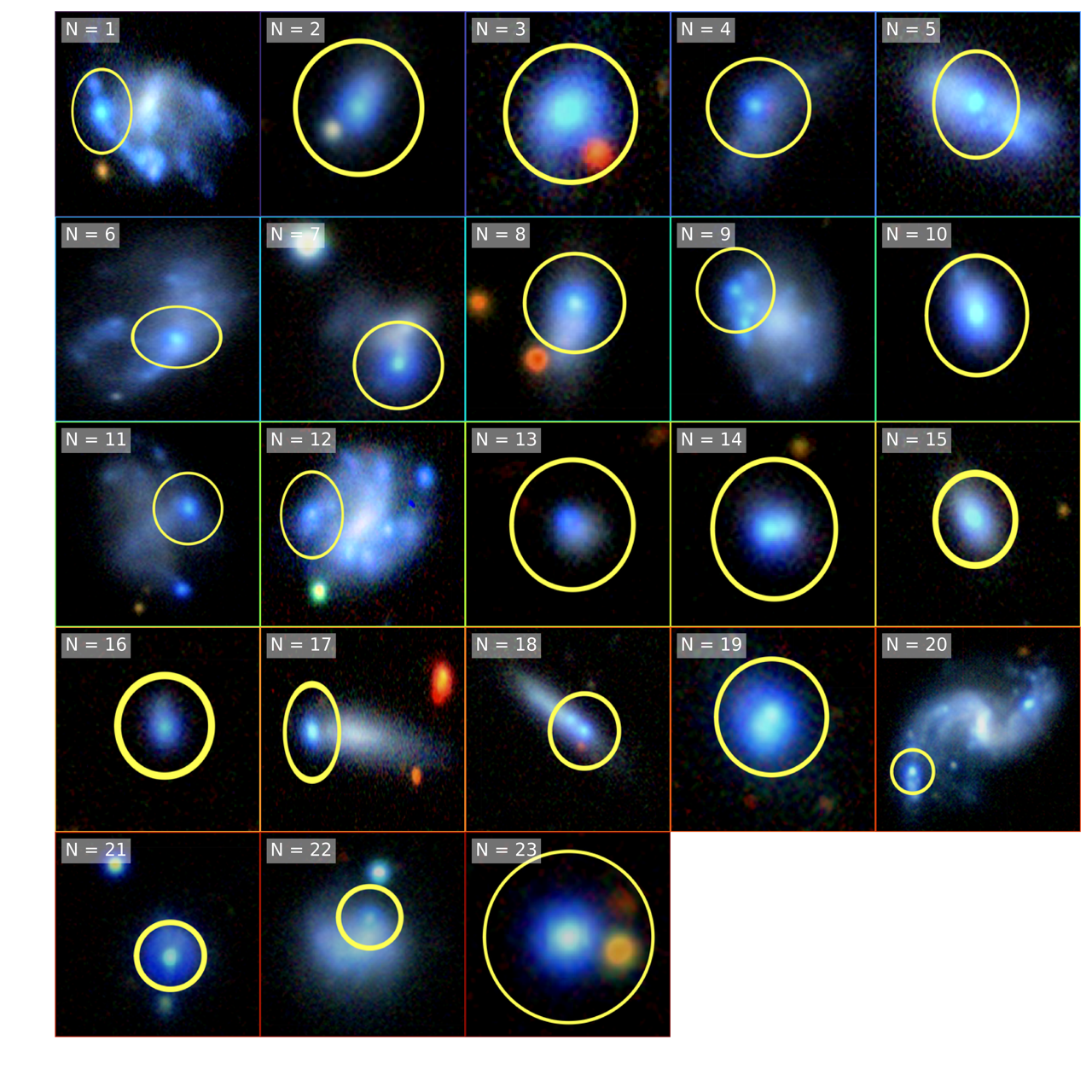}
\caption{Gallery of optical images for the DESI EELG sample with stellar masses satisfying $M_* \geq 10^{7}M_{\odot}$.}
\label{fig:Galaxy_Images_DESI}
\end{figure*}

\bibliography{main}

\end{document}